%% file: main.tex
\documentclass[10pt,journal,twoside,compsoc]{IEEEtran}

\usepackage{amsmath}
\usepackage{graphicx}
\usepackage{txfonts}
\usepackage{multirow}
\usepackage{verbatim}
\usepackage{array}
\usepackage{subfig}
\usepackage[usenames, dvipsnames]{xcolor}
\usepackage[nocompress]{cite}
\usepackage[colorlinks, linkcolor=black, urlcolor = blue, citecolor=black]{hyperref}

\usepackage{pifont}

\usepackage{soul}

\usepackage{multirow}  
\newtheorem{definition}{Definition}   

\usepackage{amsmath}
\usepackage{arydshln} 
\usepackage{color} 
\usepackage{algorithm}
\usepackage{algorithmic}

\begin{document}

\title{A Survey of Utility-Oriented Pattern Mining}

\author{Wensheng Gan,
	Jerry Chun-Wei Lin,~\IEEEmembership{Senior Member,~IEEE,}
	Philippe Fournier-Viger,\\
	Han-Chieh Chao, 
	Vincent S. Tseng,~\IEEEmembership{Senior Member,~IEEE,}
	and Philip S. Yu,~\IEEEmembership{Fellow,~IEEE}
	\IEEEcompsocitemizethanks{
		\IEEEcompsocthanksitem Wensheng Gan is with Harbin Institute of Technology (Shenzhen), Shenzhen, China, and with University of Illinois at Chicago, IL, USA. Email: wsgan001@gmail.com
		
		\IEEEcompsocthanksitem Jerry Chun-Wei Lin is with the Western Norway University of Applied Sciences, Bergen, Norway. Email: jerrylin@ieee.org

		\IEEEcompsocthanksitem Philippe Fournier-Viger is with Harbin Institute of Technology (Shenzhen), Shenzhen, China. Email: philfv8@yahoo.com
				
		\IEEEcompsocthanksitem Han-Chieh Chao is with the National Dong Hwa University, Hualien, Taiwan, R.O.C. Email: hcc@ndhu.edu.tw

		\IEEEcompsocthanksitem Vincent S. Tseng is with Department of Computer Science of National Chiao Tung University, Taiwan, R.O.C. Email: vtseng@cs.nctu.edu.tw
		
		\IEEEcompsocthanksitem Philip S. Yu is with University of Illinois at Chicago, IL, USA. Email: psyu@uic.edu
	
	Manuscript received May 2018; revised May 2019 and Aug. 2019; accepted Sept. 2019. Date of publication XX 2019; date of current version XX 2019.  (Corresponding author: Jerry Chun-Wei Lin)

	}

}

\IEEEtitleabstractindextext{%

\begin{abstract}
 
The main purpose of data mining and analytics is to find novel, potentially useful patterns that can be utilized in real-world applications to derive beneficial knowledge. For identifying and evaluating the usefulness of different kinds of patterns, many techniques and constraints have been proposed, such as support, confidence, sequence order, and utility parameters (e.g., weight, price, profit, quantity, satisfaction, etc.). In recent years, there has been an increasing demand for utility-oriented pattern mining (UPM, or called utility mining). UPM is a vital task, with numerous high-impact applications, including cross-marketing, e-commerce, finance, medical, and biomedical applications. This survey aims to provide a general, comprehensive, and structured overview of the state-of-the-art methods of UPM. First, we introduce an in-depth understanding of UPM, including concepts, examples, and comparisons with related concepts. A taxonomy of the most common and state-of-the-art approaches for mining different kinds of high-utility patterns is presented in detail, including Apriori-based, tree-based, projection-based, vertical-/horizontal-data-format-based, and other hybrid approaches. A comprehensive review of advanced topics of existing high-utility pattern mining techniques is offered, with a discussion of their pros and cons. Finally, we present several well-known open-source software packages for UPM. We conclude our survey with a discussion on open and practical challenges in this field.

\end{abstract}


}

\maketitle

\input{1.intro.tex}
\input{2.basicconcept.tex}

\input{3.basicupm.tex}

\input{4.advanced.tex}

\input{5.challenges.tex}
\input{6.conclusion}

\bibliographystyle{IEEEtran}
\bibliography{main}



\vspace{-1.3cm}
\begin{IEEEbiography}[{\includegraphics[width=1in,height=1.25in,clip,keepaspectratio]{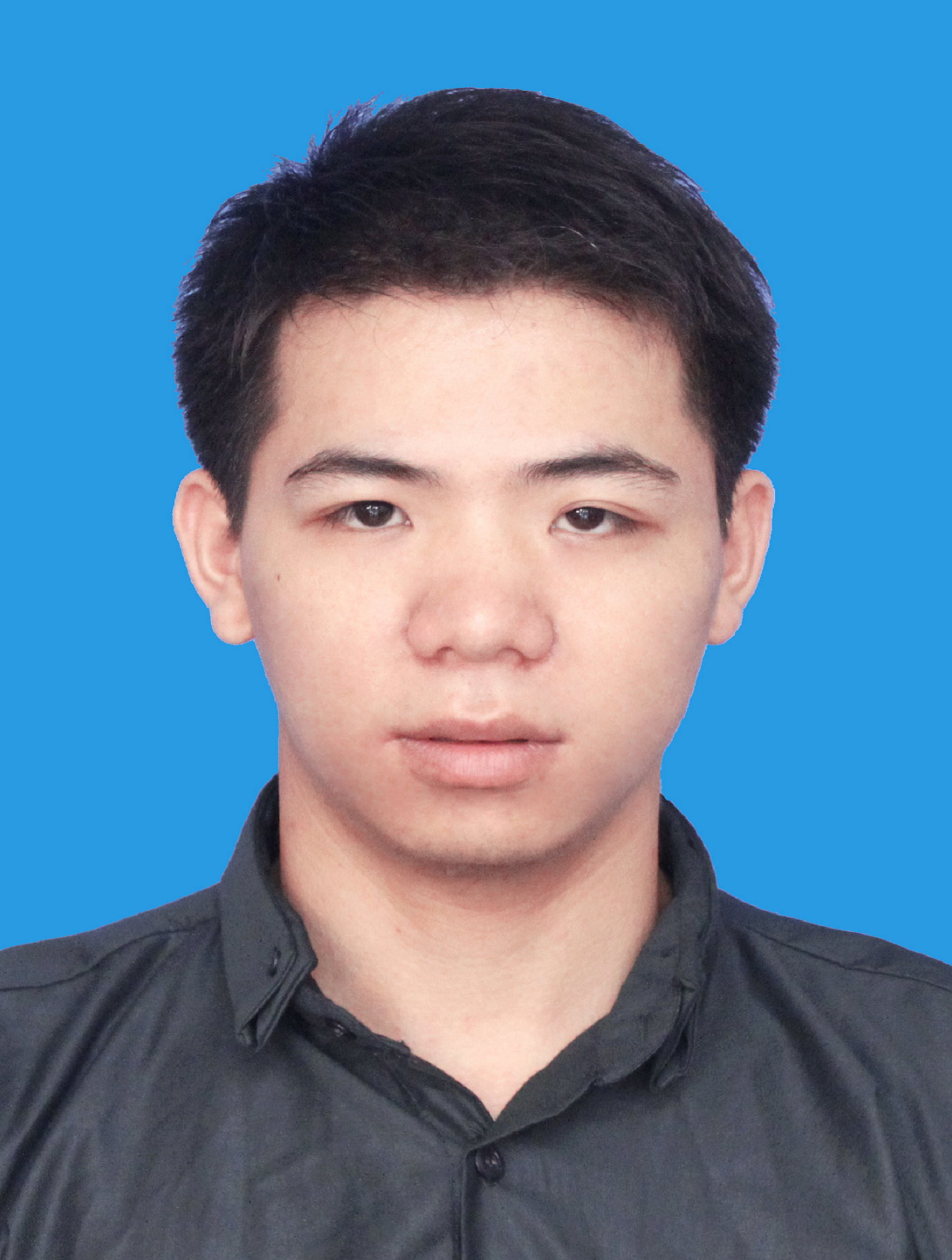}}]{Wensheng Gan} received the Ph.D. in Computer Science and Technology, Harbin Institute of Technology (Shenzhen), Guangdong, China in 2020. He wa a joint PhD student at the University of Illinois at Chicago (UIC), USA, from 2017 to 2019. He received the B.S. degree in Computer Science from South China Normal University, Guangdong, China in 2013. His research interests include data mining, utility computing, and big data analytics. He has published more than 50 research papers in  peer-reviewed journals (i.e., TKDE, TKDD, TCYB, AEI, KBS) and conferences, which have received more than 600 citations.
\end{IEEEbiography}

\vspace{-1.3cm}
\begin{IEEEbiography}[{\includegraphics[width=1in,height=1.25in,clip,keepaspectratio]{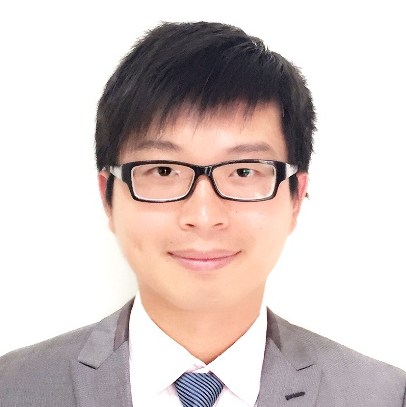}}]{Jerry Chun-Wei Lin (SM'19)}
	is an associate professor at Western Norway University of Applied Sciences, Bergen, Norway. He received the Ph.D. in Computer Science and Information Engineering, National Cheng Kung University, Tainan, Taiwan in 2010. His research interests include data mining, big data analytics, and social network. He has published more than 300 research papers in peer-reviewed international conferences and journals, which have received more than 3200 citations. He is the co-leader of the popular SPMF open-source data mining library and the Editor-in-Chief (EiC) of the \textit{Data Mining and Pattern Recognition} (DSPR) journal, and Associate Editor of \textit{Journal of Internet Technology}. 
\end{IEEEbiography}

\vspace{-1.3cm}
\begin{IEEEbiography}[{\includegraphics[width=1in,height=1.25in,clip,keepaspectratio]{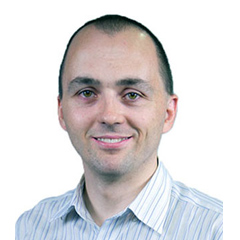}}]{Philippe Fournier-Viger}
	is full professor and Youth 1000 scholar at the Harbin Institute of Technology (Shenzhen), Shenzhen, China. He received a Ph.D. in  Computer Science at the University of Quebec in Montreal (2010). His research interests include pattern mining, sequence analysis and prediction, and social network mining. He has published more than 250 research papers in refereed international conferences and journals. He is the founder of the popular SPMF open-source data mining library, which has been cited in more than 800 research papers. He is Editor-in-Chief (EiC) of the \textit{Data Mining and Pattern Recognition} (DSPR) journal.
\end{IEEEbiography}

\vspace{-1.3cm}
\begin{IEEEbiography}[{\includegraphics[width=1in,height=1.25in,clip,keepaspectratio]{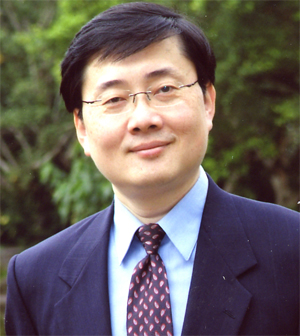}}]{Han-Chieh Chao (SM'04)}
	has been the president of National Dong Hwa University since February 2016. He received M.S. and Ph.D. degrees in Electrical Engineering from Purdue University in 1989 and 1993, respectively. His research interests include high-speed networks, wireless networks, IPv6-based networks, and artificial intelligence. He has published nearly 500 peer-reviewed research papers. He is the Editor-in-Chief (EiC) of IET Networks and \textit{Journal of Internet Technology}. Dr. Chao has served as a guest editor for ACM MONET, IEEE JSAC, \textit{IEEE Communications Magazine}, \textit{IEEE Systems Journal}, \textit{Computer Communications}, \textit{IEEE Proceedings Communications}, \textit{Wireless Personal Communications}, and \textit{Wireless Communications \& Mobile Computing}. Dr. Chao is an IEEE Senior Member and a fellow of IET. 
\end{IEEEbiography}

\vspace{-1.3cm}
\begin{IEEEbiography}[{\includegraphics[width=1in,height=1.25in,clip,keepaspectratio]{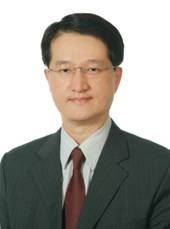}}]{Vincent S. Tseng (SM'16)}
	is currently a Distinguished Professor at the Department of Computer Science of National Chiao Tung University, Taiwan.  Dr. Tseng received his Ph.D. degree with a major in computer science from National Chiao Tung University, in 1997. His research interests covering data mining, Big Data, biomedical informatics, mobile and Web technologies. He has published more than 400 research papers in peer-reviewed journals and conferences and holds 15 patents. He has been on the editorial board of a number of journals, including \textit{IEEE Transactions on Knowledge and Data Engineering}, and \textit{ACM Transactions on Knowledge Discovery from Data}, \textit{IEEE Journal of Biomedical and Health Informatics}.  Dr. Tseng is an IEEE senior member.
\end{IEEEbiography}

\vspace{-1.3cm}
\begin{IEEEbiography}[{\includegraphics[width=1in,height=1.25in,clip,keepaspectratio]{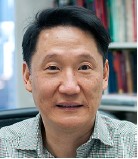}}]{Philip S. Yu (F'93)}
	received the B.S. degree in electrical engineering from National Taiwan University, M.S. and Ph.D. degrees in electrical engineering from Stanford University, and an MBA from New York University. He is a distinguished professor of computer science with the University of Illinois at Chicago (UIC) and holds the Wexler Chair in Information Technology at UIC. Before joining UIC, he was with IBM, where he was manager of the Software Tools and Techniques Department at the Thomas J. Watson Research Center. His research interests include databases, data mining,  artificial intelligence, and privacy. He has published more than 1,300 papers in peer-reviewed journals (i.e., TKDE, TPDS, TKDD, VLDBJ) and conferences (i.e., SIGMOD, KDD, ICDE, WWW, AAAI, SIGIR, ICML, etc). He holds or has applied for more than 300 U.S. patents. Dr. Yu was the Editor-in-Chief of \textit{ACM Transactions on Knowledge Discovery from Data}. He received the ACM SIGKDD 2016 Innovation Award, and the IEEE Computer Society 2013 Technical Achievement Award. Dr. Yu is a fellow of the ACM and the IEEE.
\end{IEEEbiography}

\end{document}

%% file: 1.intro.tex
\section{Introduction}
\label{sec:introduction} 

\IEEEPARstart{D}{ata} mining \cite{chen1996data,han2011data} focuses on extraction of information from a large set of data and transforms it into an easily interpretable structure for further use. It is an interdisciplinary field focused on scientific methods, processes, and systems to extract knowledge or insights from data in various forms, either structured or unstructured.  Mining interesting patterns from different types of data is quite important in many real-life applications \cite{chen1996data,koh2016unsupervised,fournier2017survey0,fournier2017survey,tsai2014data}. In recent decades, the task of interesting pattern mining [e.g., \textit{frequent pattern mining} (FPM) \cite{han2004mining,aggarwal2009frequent}, \textit{association rule mining} (ARM) \cite{agrawal1993mining,agrawal1994fast},  \textit{frequent episode mining} (FEM) \cite{mannila1997discovery,huang2008efficient,achar2012unified,achar2013pattern}, and \textit{sequential pattern mining} (SPM) \cite{agrawal1995mining,han2001prefixspan,fournier2017survey,gan2019survey}] has been extensively studied. These are important and fundamental data mining techniques \cite{chen1996data} that satisfy the requirements of real-world applications in numerous domains. Most of them aim at extracting the desired patterns using frequency or co-occurrence \cite{agrawal1993mining,agrawal1994fast,han2004mining,aggarwal2009frequent}, as well as other properties and interestingness measures \cite{geng2006interestingness,pei2001mining,tan2004selecting,mcgarry2005survey}.  Despite the wide use of pattern mining techniques, most of these algorithms do not allow for the discovery of utility-oriented patterns, i.e., those that contribute the most to a predefined utility threshold, an objective function, or a performance metric. In general, some implicit factors, such as the utility, interestingness, or risk of objects/patterns, are commonly seen in real-world situations. The knowledge that is actually important to the user may not be found by traditional data mining algorithms. Therefore, a novel utility mining framework, called \textit{utility-oriented pattern mining (UPM)} or \textit{high-utility pattern mining (HUPM\footnote{The terms of UPM and HUPM can be interchangeably used but we will use UPM in the rest of this manuscript.})} \cite{shen2002objective,chan2003mining,yao2006unified}, which considers the relative importance of items (\textit{utility-oriented} \cite{marshall1926principles}), has become an emerging research topic in recent years. In UPM, the \textit{utility} (i.e., importance, interest, satisfaction, or risk) of each item can be predefined based on a user's background knowledge or preferences. 

According to Wikipedia\footnote{\url{https://en.wikipedia.org/wiki/Utility}}, in economics, utility is a measure of preferences over some set of goods (including services, i.e., something that satisfies human wants); it represents satisfaction experienced by the consumer of a good. Hence, utility is a subjective measure. This definition indicates that a subjective value is associated with a specific value in a domain to express user preference. In practice, the value of utility is assigned by the user according to his interpretation of domain-specific knowledge measured by a specific value, such as cost, profit, or aesthetic value. According to the studies of Li \textit{et al}. \cite{geng2006interestingness}, interestingness measures can be classified as objective measures, subjective measures, and semantic measures \cite{tan2004selecting,geng2006interestingness,mcgarry2005survey}. Objective measures \cite{hilderman2003measuring,mcgarry2005survey}, such as \textit{support} or \textit{confidence} for pattern mining, are based only on data itself, whereas subjective measures \cite{silberschatz1995subjective,de2011maximum}, such as \textit{unexpectedness} or \textit{novelty}, take into account the user's domain knowledge. For the semantic measures \cite{yao2006unified}, such as utility, they consider the data itself, as well as the user's expectation. Hence, utility is a quantitative representation of user preference, and the usefulness of an itemset is quantified in terms of its utility value. Utility can be defined as ``A measure of how `useful' (i.e., profitable) an itemset is" \cite{yao2006unified,yao2006mining}. Formally, a pattern is said to be useful to a user if it satisfies a specific utility constraint. In practice, the utility value of a pattern can be measured in terms of cost, profit, aesthetic value, or other measures of user preference.

\begin{figure}[!htbp]
	\setlength{\abovecaptionskip}{0pt}
	\setlength{\belowcaptionskip}{0pt}	
	\centering
	\includegraphics[width=3.5in]{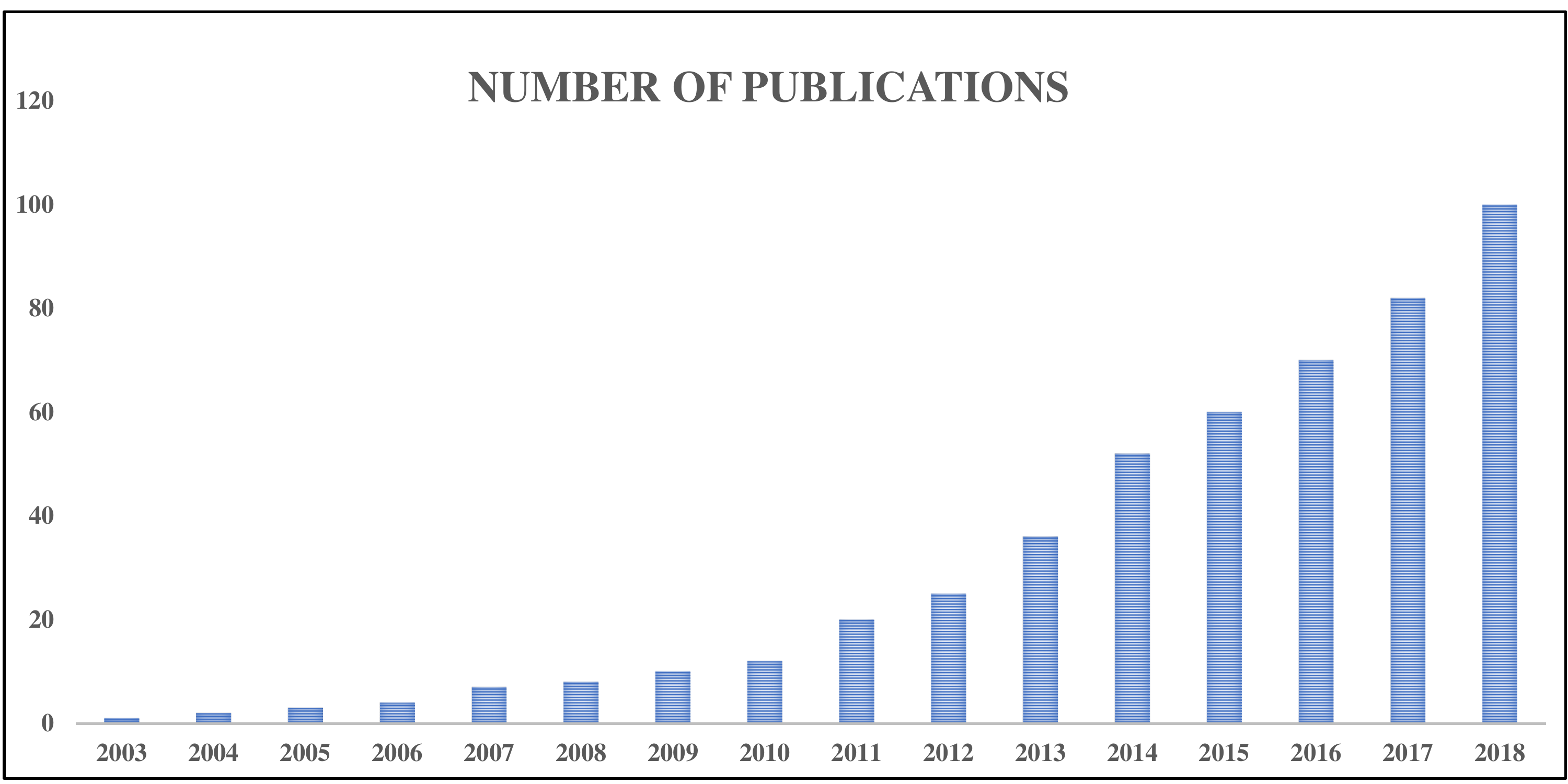}
	\caption{Number of published papers that use ``High Utility Pattern Mining" in sub-areas of data science and analytics. These publication statistics are obtained from \textit{Google Scholar}. Note that the search phrase is defined as the sub-field named with the exact phrase ``utility pattern," and at least one of utility or itemset/rule/episode/sequential pattern appearing, e.g., ``utility itemset," ``utility pattern," ``utility sequential pattern".}
	
	\label{fig:PublicationOfHUPM}
\end{figure}

To address these issues, \textit{utility-oriented pattern mining} (hereinafter called UPM) has become a useful task and an important topic in data mining. In UPM, each object/item has an unit \textit{utility} (e.g., unit profit) and can appear more than once in each transaction or event (e.g., purchase quantity). The \textit{utility} of a pattern represents its importance or satisfaction, which can be measured in terms of risk, profit, cost, quantity, or other information depending on user preference. In general, the utility of a pattern is based on local transaction utility (also called \textit{internal utility}) and \textit{external utility} \cite{yao2006unified,yao2006mining}. The \textit{internal utility} of an object/item is defined according to the information stored in a transaction/event, such as the quantity of the object/item occurred or sold.  The \textit{external utility} can be a measure for describing user preferences. Therefore, the utility of a pattern depends on the \textit{utility function} specified by the user, which can be the \textit{Sum}, \textit{Average}, or \textit{Multiplication} of quantity and profit of this pattern in databases. More specifically, the utility-based method for pattern mining can find various types of patterns that could not be identified using previous theories and techniques. According to previous studies, UPM has a wide range of applications, including website click-stream analysis \cite{li2008fast,ahmed2009efficient,shie2010online}, cross-marketing in retail stores \cite{erwin2008efficient,li2008isolated}, mobile commerce environment \cite{shie2011mining,shie2013efficient}, gene regulation \cite{zihayat2017mining}, and biomedical applications \cite{yen2007mining}. Through 15 years of study and development, many techniques and approaches have been extensively proposed for UPM in various applications. As shown in Fig. \ref{fig:PublicationOfHUPM}, there has been a rapid surge of interest of UPM in recent years in terms of the number of academic  papers published in several sub-fields, including high-utility itemsets \cite{yao2006mining}, high-utility rules \cite{lee2013utility,sahoo2015efficient}, high-utility sequential patterns \cite{ahmed2010novel,yin2012uspan}, and high-utility episodes \cite{wu2013mining,lin2015discovering}.

In spite of the fact that there are a considerable number of existing published studies and surveys about data mining, especially for pattern mining, none of them discuss UPM. Yet, after more than 15 years of theoretical development, a significant number of new technologies and applications have appeared in the UPM field. Unfortunately, there is no comprehensive survey of utility-oriented pattern mining methods and no study that systematically compares the state-of-the-art algorithms. We believe that now is a good time to summarize the new technologies and address the gap between theory and application. Here, we attempt to find a clearer way to present the concepts and practical aspects of UPM for the data mining research community. In this paper, we provide a systematic and comprehensive survey of the significant advances in UPM. The methods discussed in this article are not only important for high-utility pattern (i.e., itemset \cite{yao2006unified,yao2006mining}, rule \cite{lee2013utility,sahoo2015efficient}, sequence, episode, etc.) mining but can also serve as inspiration for other data mining tasks \cite{chen1996data,han2011data}, including episode mining \cite{mannila1997discovery,huang2008efficient,achar2012unified,achar2013pattern}, distributed data mining \cite{gan2017data}, and incremental/dynamic data mining \cite{nath2013incremental,gan2018survey}.  The major contributions are listed as follows:

\begin{enumerate}
	\item This paper first presents the background, motivation, and a comprehensive survey of UPM (Section\ref{sec:introduction}). This survey investigates more than 150 UPM papers published in the last 15 years and summarizes them in a systematic fashion. 
	
	\item  This survey first introduces an in-depth understanding of UPM, including concepts, examples, comparisons with related studies (e.g., FPM, SPM), applications, and evaluation measures (Section \ref{sec:concepts}). This survey presents a bird's eyes view, and then deeply and comprehensively summarizes the developments of UPM, comparing the state-of-the-art works to earlier works (Section \ref{sec:basicapproaches}). 
	
	\item A taxonomy of the most common and the state-of-the-art approaches for UPM is presented, including Apriori-based, tree-based, projection-based, vertical/horizontal-data-format-based, and other hybrid approaches (Section \ref{sec:basicapproaches}). We further analyze the pros and cons of each presented approach.
	
	\item A comprehensive review of advanced topics of utility  mining techniques (e.g., dynamic UPM, concise representation of utility patterns, HUSPM, HUEM, UPM in big data, and privacy preserving for UPM) is presented (Section \ref{sec:advancedtopic}), with a discussion of their pros and cons. Not only the representative algorithms but also the advances and latest progress are reviewed. 
	
	\item  We further review some well-known open-source software  and datasets (Section \ref{sec:softwaredatasets} of UPM and hope that these resources may reduce barriers for future research. Finally, we identify several important issues and research opportunities for UPM (Section \ref{sec:challenges}). 
\end{enumerate}

The remainder of this article is organized as follows. In Section \ref{sec:concepts}, we introduce the necessary background information, the basic concepts and examples, and the applications in this field. In Section \ref{sec:basicapproaches}, we give a high-level overview of emerging UPM problems and survey several popular methods, as well as recent developments. In Section \ref{sec:advancedtopic}, we discuss advanced topics and techniques of UPM. In addition, several well-known open-source software and datasets are summarized in Section \ref{sec:softwaredatasets}. We describe some open challenges and opportunities in Section \ref{sec:challenges}. Several future directions are described in Section \ref{sec:conclusion}.

%% file: 2.basicconcept.tex
\section{Basic Concept: Utility-Oriented Pattern Mining}
\label{sec:concepts}

\begin{table}[!htbp]
	\centering
	\scriptsize 
	\caption{Summary of symbols and their explanations}
	\label{table_Notation}
	\begin{tabular}{|c|l|}
		\hline
		\textbf{Symbol} & \multicolumn{1}{c|}{\textbf{Definition}} \\ \hline
		$I$ &  A set of $m$ distinct items, \textit{I} = \{\textit{i}$_{1}$, \textit{i}$_{2}$, $\ldots$, \textit{i$_{m}$}\}. \\ \hline

    	$D$	&  A quantitative database, \textit{D} = \{\textit{T}$_{1}$, \textit{T}$_{2}$, $\ldots$, \textit{T$_{n}$}\}. \\ \hline

		\textit{QSD} &  A quantitative sequential database = \{\textit{s}$_{1}$, \textit{s}$_{2}$, $\ldots$, \textit{s$_{n}$}\}.  \\ \hline

		$k$-itemset &   An itemset with $k$ number of items in itself. \\ \hline
				
		$X$ &   A $k$-itemset having $k$ distinct items $\{i_1, i_2, \ldots, i_k\}$.  \\ \hline

		$sup(X)$ &   The support of a pattern $X$ in \textit{D} or \textit{QSD}.  \\ \hline	

		$q(i_j, T_q)$ &   The purchase quantity of an item $i_j$ in transaction $T_q$. \\ \hline
		
		$pr(i_j)$ &   The predefined unit profit of an item $i_j$. \\ \hline
				
		$u(i_j, T_q)$ &  The utility of an item $i_j$ in transaction $T_q$. \\ \hline
						
		$u(X, T_q)$ &  The utility of an itemset $X$ in transaction $T_q$. \\ \hline

		$tu(T_q)$ &   The sum of the utilities of items in transaction $T_q$. \\ \hline

		$minsup$ &   A predefined minimum support threshold. \\ \hline

		$minconf$ &   A predefined minimum confidence threshold. \\ \hline
				
		$minutil$ &   A predefined minimum high-utility threshold. \\ \hline

		$TWU$ &   The transaction-weighted utility of a pattern. \\ \hline
		
		\textit{TWDC} &   Transaction-weighted downward closure property. \\ \hline
		
		\textit{HTWUI} &   A high transaction-weighted utilization itemset. \\ \hline		
		\textit{HUI} &   A high-utility itemset. \\ \hline

		\textit{FPM} &   Frequent pattern mining. \\ \hline
		\textit{SPM} &   Sequential pattern mining. \\ \hline
		\textit{UPM} &   Utility-oriented pattern mining. \\ \hline

		\textit{HUAR} &   High-utility association rule. \\ \hline
		\textit{HUSP} &   High-utility sequential pattern. \\ \hline
		\textit{HUSR} &   High-utility sequential rule. \\ \hline
		\textit{HUE} &   High-utility episode. \\ \hline
		\textit{HUIM} &   High-utility itemset mining. \\ \hline
		\textit{HUARM} &   High-utility association rule mining. \\ \hline		
		\textit{HUSPM} &   High-utility sequential pattern mining. \\ \hline		
		\textit{HUEM} &   High-utility episode mining. \\ \hline
	\end{tabular}
\end{table}

\subsection{Preliminary and Types of Utility-Oriented Patterns}
We first present the basic notations, as summarized in Table \ref{table_Notation}. Then, we introduce related preliminaries of UPM and then define the problem of UPM.  Based on pattern diversity, utility-oriented pattern mining can be classified using the following basic criteria and extended patterns.

\begin{definition}[frequent pattern and association rule \cite{agrawal1993mining}]
	\rm An association rule is an \textit{implication} of the form, $X \rightarrow Y$, where $X \subset I$, $Y \subset I,$ and $X \cap Y = \phi$. $X$ (or $Y$) is a set of \textit{items}, called an \textit{itemset}.  Given a database, a pattern (e.g., a set of items, sequences, structures, etc.) is said to be a frequent pattern if it occurs frequently in this database. Support of an itemset denoted as $sup(X)$ is the number of transactions containing $X$. An association rule $R$: $X \rightarrow Y$, in which $X$, $Y$ are disjoint, and $Y$ is non-empty, means that if a transaction includes $X$, then it also has $Y$. An association rule consists of frequent itemsets, and its confidence is no less than the \textit{minimum confidence} sometimes called \textit{strong rules}. It was first proposed by Agrawal et al. \cite{agrawal1993mining} in the context of frequent itemset and association rule mining. For example, $\{Cheese, Milk\}$ $ \rightarrow $ \textit{Bread} \textit{[sup = 5\%, conf = 80\%]}; this association rule means that 80\% of customers who buy \textit{Cheese} and \textit{Milk} also buy \textit{Bread}, and 5\% of customers buy all these products together.
\end{definition}

\begin{definition}[high-utility itemset, HUI \cite{yao2006mining,tseng2013efficient}]
	\rm The utility of an item $ i_{j} $ appearing in a transaction $ T_{q} $ is denoted as $u(i_{j}, T_{q}) $ and defined as $u(i_{j}, T_{q})$ = $q(i_{j}, T_{q})\times pr(i_{j})$. The utility of an itemset \textit{X} in $ T_{q} $ is defined as $u(X, T_{q})$ = $\sum _{i_{j}\in X\wedge X\subseteq T_{q}}u(i_{j}, T_{q})$. The total utility of \textit{X} in a database \textit{D} is $u(X)$ = $\sum_{X\subseteq T_{q}\wedge T_{q}\in D} u(X, T_{q})$. An itemset is said to be a \textit{high-utility itemset} (HUI) if its total utility in a database is no less than the user-specified minimum utility threshold (such that $u(X) \geq minutil $); otherwise, it is called a \textit{low-utility itemset}.
\end{definition}

\begin{definition}[high-utility association rule, HUAR \cite{lee2013utility,sahoo2015efficient}]
	\rm Since the usefulness of association rule \cite{agrawal1993mining} can be defined as a \textit{utility function} based on the business objective, the \textit{utility} and \textit{confidence} can be used to extend the concepts of high-utility itemset and association rule. An association rule $R$: $X \rightarrow Y$ is considered to have high utility if it meets the \textit{minutil} constraint. Thus, a \textit{high-utility association rule} (HUAR) consists of high-utility itemsets, and its confidence is no less than the \textit{minimum confidence}. Generally speaking, discovery of HUIs started as the first phase in the discovery of HUARs, but it has been generalized by formulating a new pattern-mining framework.
\end{definition}

\begin{definition}[high-utility sequential pattern, HUSP \cite{ahmed2010novel,yin2012uspan}]
	\rm The utility of an item ($ i_{j} $) in a \textit{q}-itemset \textit{v} is denoted as $ u(i_{j}, v) $, and defined as	$u(i_{j}, v)$ = $q(i_{j}, v)\times pr(i_{j}),$ where $ q(i_{j}, v) $ is the quantity of ($ i_{j} $) in $ v $, and $ pr(i_{j}) $ is the profit of ($ i_{j} $). The utility of a \textit{q}-itemset $ v $ is denoted as $ u(v) $ and defined as	$u(v)$ = $\sum_{i_{j}\in v}u(i_{j}, v)$. The utility of a \textit{q}-sequence $ s $ = $<$$v_{1}, v_{2}$, $\dots, v_{d}$$>$ is denoted as $ u(s) $ and defined as $u(s)$ = $\sum_{v\in s}u(v)$.  A sequence $ s $ in a quantitative sequential database \textit{QSD} is said to be a \textit{high-utility sequential pattern} (HUSP) if its utility is no less than the minimum threshold of $ s $ as $HUSP$  $\gets\{s|u(s) \geq minutil\}$. Considering the ordered sequences, high-utility sequential pattern mining (HUSPM) \cite{ahmed2010novel,yin2012uspan} can discover more informative sequential patterns. This process is more complicated than the traditional UPM or SPM since the order and the utilities of itemsets should be considered together.

\end{definition}

\begin{definition}[high-utility sequential rule, HUSR \cite{zida2015efficient}]
	\rm   A sequential rule $R$: $X \rightarrow Y$ \cite{fournier2012cmrules} is a relationship between two unordered itemsets $X$, $Y \subseteq I $ such that $X$ $\cap$ $Y$ = $\emptyset$ and $X$, $Y$ $\neq \emptyset$. The interpretation of a rule $R$: $X \rightarrow Y $ is that if items of $X$ occur in a sequence, then items of $Y$ will occur afterward in the same sequence. Let \textit{minsup}, \textit{minconf} $ \in $ [0, 1] and  \textit{minutil}  be thresholds set by the user and \textit{QSD} be a sequence database. A sequential rule $R$ is said to be a \textit{high-utility sequential rule} (HUSR) \cite{zida2015efficient} iff $u(R) \geq minutil$ and $R$ is a valid rule, in which $u(R)$ is the total utility of $R$ in \textit{QSD}. Otherwise, it is said to be a \textit{low-utility sequential rule}. The problem of mining high-utility sequential rules from a sequence database is the discovery of all high-utility sequential rules. 
\end{definition}

\begin{definition}[high-utility episode, HUE \cite{wu2013mining,lin2015discovering}]
	\rm  An \textit{episode} $\alpha$ is a non-empty totally ordered set of  \textit{simultaneous events} (\textit{SE}) of the form $<$$(SE_1), (SE_2), \dots, (SE_k)$$>$, where $SE_i$ appears before $SE_j$ for all 1 $ \leq i < j \leq k$. For example, $<$$(AB), (C)$$>$ is an episode containing a simultaneous event ($AB$) and a series event ($C$). The total utility of an episode $\alpha$ in a single simple or complex event containing a set of sub-events is $u(\alpha)$ \cite{wu2013mining,lin2015discovering}, and its calculation is more complicated than that of the utility of a sequence \cite{yin2012uspan}. An episode is said to be a \textit{high-utility episode} (abbreviated as HUE) in complex event sequences if its total utility in these sequences is no less than the minimum utility threshold such that $u(\alpha) \geq minutil$. Otherwise, this episode is a \textit{low-utility episode}. 

\end{definition}

\begin{definition}[utility-oriented pattern mining, UPM]
	\rm   A general definition of UPM  is given below: UPM is a new mining framework that utilizes the utility theory and various mining techniques (e.g., data structure, pruning strategy, upper bound) to discover the interesting patterns (e.g., HUI, HUAR, HUSP, HUSR, HUE), and these derived patterns can lead to utility maximization and high benefit in business or other tasks.
\end{definition}

Based on the above concepts of utility pattern, the UPM framework can be further classified into the following categories, including 1) \textit{high-utility itemset mining } (\textbf{HUIM}), 2) \textit{high-utility association rule mining} (\textbf{HUARM}), 3) \textit{high-utility sequential pattern mining} (\textbf{HUSPM}), 4) \textit{high-utility sequential rule mining} (\textbf{HUSRM}), and 5) \textit{high-utility episode mining} (\textbf{HUEM}).

\subsection{Comparisons with Related Concepts}
With the boom in data mining and analysis, all kinds of data have emerged, and a number of concepts (e.g., FPM, SPM, FEM, UPM, etc.) to model various types of data have been proposed. These concepts have similar meanings, as well as subtle differences. Here we compare the UPM framework with its most related concepts.

$\bullet$  \textbf{UPM vs. FPM}. Frequent pattern mining (FPM) \cite{agrawal1993mining,agrawal1994fast,han2004mining,aggarwal2009frequent} is a common and fundamental topic in data mining. FPM is a key phase of association-rule mining (ARM), but it has been generalized to many kinds of patterns, such as frequent sequential patterns \cite{han2001prefixspan}, frequent episodes \cite{mannila1997discovery}, and frequent subgraphs \cite{jiang2013survey}. The goal of FPM is to discover all the desired patterns having support no lower than a given \textit{minimum support} threshold. If a pattern has higher support than this threshold, it is called a frequent pattern; otherwise, it is called an infrequent pattern. Unlike UPM,  studies of FPM seldom consider the database having quantities of items, and none of them considers the utility feature. Under the ``economic view" of consumer rational choices, utility theory can be used to maximize the estimated profit. UPM considers both statistical significance and profit significance, whereas FPM aims at discovering the interesting patterns that frequently co-occur in databases. In other words, any frequent pattern is treated as a significant one in FPM. However, in practice, these frequent patterns do not show the business value and impact. In contrast, the goal of UPM is to identify the useful patterns that appear together and also bring high profits to the merchants \cite{yao2004foundational}. In UPM, managers can investigate the historical databases and extract the set of patterns having high combined utilities. Such problems cannot be tackled by the support/frequency-based FPM framework.

$\bullet$  \textbf{UPM vs. WFPM}. In the related areas, the relative importance of each object/item is not considered in the concept of FPM. To address this problem, weighted frequent-pattern mining (WFPM) was proposed \cite{cai1998mining,wang2000efficient,tao2003weighted,sun2008mining,lin2015rwfim,lin2016weighted,gan2017extracting1}. In the framework of WFPM, the weights of items, such as unit profits of items in transaction databases, are considered. Therefore, even if some patterns are infrequent, they might still be discovered if they have high \textit{weighted support} \cite{cai1998mining,wang2000efficient,tao2003weighted}. However, the quantities of objects/items are not considered in WFPM. Thus, the requirements of users who are interested in discovering the desired patterns with high risks or profits cannot be satisfied. The reason is that the profits are composed of unit profits (i.e., weights) and purchased quantities. In view of this, utility-oriented pattern mining has emerged as an important topic. It refers to discovering the patterns with high profits. As mentioned previously, the meaning of a pattern's utility is the interestingness, importance, or profitability of the pattern to users. The utility theory is applied to data mining by considering both the unit utility (i.e., profit, risk, and weight) and purchased quantities. This has led to the concept of UPM \cite{yao2004foundational}, which selects interesting patterns based on \textit{minimum utility} rather than \textit{minimum support}.

$\bullet$  \textbf{UPM vs. SPM}. Different from FIM, sequential pattern mining (SPM) \cite{agrawal1995mining,han2001prefixspan,fournier2017survey,gan2019survey}, which discovers frequent subsequences as patterns in a sequence database that contains the embedded timestamp information of an event, is more complex and challenging. In 1995, Agrawal and Srikant first extended the FPM model to handle sequences \cite{agrawal1995mining}. Consider the sequence $<$$\{a,e\}, \{b\}, \{c,d\}, \{g\}, \{e\}$$>$, which represents five events made by a customer at a retail store. Each single letter represents an item (i.e., $\{a\}$, $\{c\}$, $\{g\}$, etc.), and items between curly braces represent an itemset (i.e., $\{a,e\}$ and $\{c,d\}$). Simply speaking, a sequence is a list of temporally ordered itemsets (also called events). Owing to the absence of time constraints in FPM not present in SPM, SPM has a potentially huge set of candidate sequences \cite{han2001prefixspan}. In a related area, through 25 years' study and development, many techniques and approaches have been proposed for mining sequential patterns in a wide range of real-world applications \cite{fournier2017survey}. In general, SPM mainly focuses on the co-occurrence of derived patterns; it does not consider the unit profit and purchase quantities of each product/item.

So far, we have reviewed a wide range of pattern-mining frameworks that aim to discover various types of patterns, such as itemsets \cite{agrawal1993mining,cai1998mining}, sequences \cite{agrawal1995mining,han2001prefixspan}, and graphs \cite{jiang2013survey}. These frameworks, however, only select high-frequency/support patterns. Patterns below the minimum threshold are considered useless and discarded. \textit{Frequency} is the main interestingness measurement, and all objects/items and transactions are treated equally in such a framework. Clearly, this assumption contradicts the truth in many real-world applications because the importance of different items/itemsets/sequences might be significantly different. In these circumstances, the frequency-/support-based framework is inadequate for pattern mining and selection. Based on the above concerns, researchers proposed the concept of UPM.

\subsection{Why Utility-Oriented Pattern Mining and Analysis}

With the rapid advancement of research on UPM, numerous applications in different domains have been proposed in recent years. We next describe several important applications, as summarized in Table \ref{table_applications}.

\begin{table}[!htbp]

	\centering
	\scriptsize  
	\caption{Various applications of UPM.}
	\label{table_applications}
	\newcommand{\tl}[1]{\multicolumn{1}{l}{#1}} 
	\begin{tabular}{|l|l|l|l|l|l|} 
		\hline
		\multicolumn{1}{|c|}{\textbf{Domain}} & \multicolumn{1}{c|}{\textbf{Applications}}  \\ \hline
		
		\multirow{2}{1.9cm}{Business intelligence} &  \multirow{2}{5.5cm}{Market basket analysis, recommendation, cross marketing, sales intelligence, and risk prediction.}\\
		    & 	   \\  \hline

		\multirow{2}{1.9cm}{Web mining}  &  \multirow{2}{5.5cm}{Users' click-stream, users' access behaviors, and traversal pattern mining.}   \\
		&   \\ \hline

		\multirow{2}{1.9cm}{Mobile computing}  & 	\multirow{2}{5.5cm}{Mobile e-commerce, travel route recommendation, spatial crowdsourcing, and spatial data analytics.}  \\
		&   \\ \hline

		\multirow{2}{1.9cm}{Stream processing}  & \multirow{1}{5.5cm}{Web-click stream analysis, IoT data  analytics, and stream mining in smart transportation.}  \\ 
		&     \\ \hline

		\multirow{1}{0.5cm}{Biomedicine}     & \multirow{1}{5.5cm}{Gene expression and gene-disease association.} \\ 
		 \hline
		
	\end{tabular}
\end{table}

$\bullet$  \textbf{Market basket analysis}. In market basket analysis, each transaction recorded with a customer contains several products/items, annotated with their purchase time, purchase quantities and the selling price. An important technique is based on the theory that if customers buy a certain set of items, customers are more (or less) likely to buy another set of items. For the problem of mining-characterized association rules from market basket data, the goal is to not only discover the buying patterns of customers but also the highly profitable patterns and customers. In some existing frameworks \cite{erwin2008efficient,li2005direct,li2008isolated,lin2016fast1,gan2017extracting}, the utility (i.e., importance, interest, or risk) of each product can be predefined based on users' background knowledge or preferences.  As a result, UPM is able to offer richly detailed information about users' purchasing behaviors.

 $\bullet$    \textbf{Web mining}. There is much rich information in web data. For example, users' click-stream and purchase behaviors are recorded in web logs. In such data, a user's click operation (there are one or many clicks in one session) and browsing time on a web page can be expressed as the \textit{internal utility} of the web page. Obviously, each web page has different importance depending on users' different preferences (i.e., \textit{external utility}). Thus, UPM technology can be used to discover utility-oriented patterns from web logs, such as high-utility access patterns \cite{ahmed2011framework2} and high-utility traversal patterns \cite{ahmed2009efficient2}. The derived results are quite useful for electronic commerce for such things as improving website services, providing some navigation suggestions for web browsing, and improving the design of web pages.

$\bullet$     \textbf{Mobile computing}. With the explosive growth of the Internet of Things (IoT)  \cite{atzori2010internet} technologies in the Big Data era, such as smart-phones, wireless networks, and GPS devices, information about users' mobile behavior (e.g., locations and payment records) can be acquired and integrated in data analytics. In this scenario, utility-oriented mining technologies can be used to discover valuable user behaviors. Shie \textit{et al}. first proposed a new framework to mine high-utility mobile sequences \cite{shie2011mining,shie2013efficient} in mobile environments. It can extract associations between customers' purchase behaviors and location trajectories. The discovered high-utility patterns can be utilized for many applications essentially to mobile e-commerce, such as location-based advertisement or recommendations, navigational services, spatial crowdsourcing, and utility-based recommendation systems.

$\bullet$     \textbf{Stream processing}. The majority of data is born as continuous streams \cite{golab2003issues,chi2004moment}: sensor events, user activity on a website, financial trades, and others; all these data are created as a series of events over time. In general, some stream data contain rich and important features that are similar to the general static data. In contrast to the support-based pattern mining technologies, the utility-oriented pattern mining technologies can be applied to extract useful pattens and knowledge from stream data, i.e., website click-streams \cite{zihayat2014mining}. Some preliminary studies have been carried out on this issue, such as \cite{li2008fast,shie2010online,shie2012efficient,ahmed2012interactive}.

 $\bullet$    \textbf{Biomedicine}. In gene expression data, each row represents a set of genes and their expression levels (i.e., internal utility) under an experimental condition. In addition, each gene has a degree of importance for biological processes (i.e., external utility). In bioinformatics, UPM technology can discover useful relationships between genes. For example, Liu \textit{et al}. \cite{liu2013mining} applied a UPM method to successfully discover interesting gene regulation patterns from a time-course of comparative gene expression data. By analyzing the discovered results, medical researchers can find new drugs for the treatment of diseases. Recently, Zihayat \textit{et al}. proposed a utility model by considering both the gene-disease association scores and their degrees of expression levels in a biological investigation \cite{zihayat2017mining}.

 $\bullet$  \textbf{Other applications}. Since the ``utility" of a pattern measures the importance of the pattern to the user (i.e., risk, weight, cost, and profit), UPM has broad real-life applications; several examples are described below. In risk prediction, the risk that events may occur is indicated by occurrence probabilities and risk. For example, the event $<$($A$, 1, 80); ($D$, 5, 15); ($E$, 3, 125); 90\%$>$ indicates that this event consists of three sub-events $\{A, D, E\}$ with occurrence frequencies \{1, 5, 3\}, while their risk \{80, 15, 125\}, respectively, has a 90\% probability of occurring. In e-commerce business, this manifests as identifying customers who visit web pages a number of times by taking pages visited as a utility parameter. In financial analysis, e.g., online banking fraud detection, the transfer of a large amount of money to an unauthorized overseas account may appear once or many times in several million transactions, yet it has a substantial business impact.

\subsection{Evaluation Measures of Utility}
Here, we briefly describe several key measures that have been used in the literature to determine utility-oriented relationships in UPM. In Section 2.1, the theoretical foundations of several UPM frameworks were analyzed.  Based on the utility theory \cite{marshall1926principles}, many evaluation measures of utility have been proposed. Some commonly used evaluation measures of the utility of a pattern in the UPM field are summarized in Table \ref{table_measure}.

\begin{table}[!htbp]

	\centering
	\scriptsize  
	\caption{Evaluation measures of utility in UPM.}
	\label{table_measure}
	\newcommand{\tl}[1]{\multicolumn{1}{l}{#1}} 
	\begin{tabular}{|l|l|l|l|l|l|} 
		\hline
		\multicolumn{1}{|c|}{\textbf{Measure}} & \multicolumn{1}{c|}{\textbf{Description}}  \\ \hline
		
		&   \\ 
		Utility   & 	\multirow{2}{6.5cm}{The commonly used utility measure in many UPM models and algorithms and its definition has been given from Definitions 2 to 7, and details can be referred to \cite{yao2004foundational,yin2012uspan}.}   \\ 
		&   \\ 
		&   \\ 
		&    \\ \hline

		&   \\ 
		Average utility  &  \multirow{2}{6.5cm}{The average utility is $au(X, T_{q}) $ = $\sum_{i_{j}\in X\wedge X\subseteq T_{q}}$$q(i_{j}, T_{q})\times pr(i_{j})/|X|$ \cite{hong2011effective}, where \textit{k} is the number of items in \textit{X}. Thus, it considers the length of pattern as a major factor.}  \\
		&   \\  
		&   \\   
		&   \\ \hline

		&   \\ 
		\multirow{2}{1.5cm}{Expected/potential utility}  & 	\multirow{2}{6.5cm}{Measures both probability and utility of a pattern in uncertain databases \cite{lin2016efficient2}; the expected support \cite{chui2007mining} is measured as \textit{expSup}($X$) = $\sum_{i=1}^{|D|} (\prod_{X_i \in X}$$p(X_i, T_q))$.}   \\
		&      \\  
		&      \\   
		&      \\ \hline

		&   \\ 
		Affinitive utility  & 	\multirow{2}{6.5cm}{The affinitive utility \cite{lin2017fdhup} of a pattern $X$ in $T_q$ is defined as $au(X,T_q)$ = $eu(X) \times af(X,T_q)$, where $ eu(X)$ = $\sum_{i_j \in X}pr(i_j) $ and $ af(X,T_q) $  denote the affinitive frequency of a pattern $X$ in  $T_q$ that is $af(X,T_q)$ = $min \{q(i_1,T_q), q(i_2,T_q),$ $..., q(i_j,T_q)\}, i_j \in X $ \cite{lin2017fdhup}.}   \\ 
		&     \\  
		&     \\  
		&     \\  
		&     \\ \hline

		&   \\ 
		Utility occupancy     & \multirow{2}{6.5cm}{It depends on the contribution of a unit item. The utility occupancy  \cite{gan2019huopm}  of a pattern $X$ in $ T_q $ and $ D $ are defined as $uo(X, T_q)$ = $u(X, T_q)$/$tu(T_q) $ and $ uo(X)$ = $\sum_{X \subseteq T_q \wedge T_q \in D}$$uo(X,T_q)/|\varGamma_X| $, respectively.}  \\ 
		&     \\  
		&     \\ 
		&     \\  
		&     \\ \hline
		
	\end{tabular}
\end{table}

The most commonly adopted evaluation measure for UPM is the general \textit{utility} concept \cite{yao2004foundational,yin2012uspan}. It is based on \textit{external utility} (e.g., profit, unit price, risk) and \textit{internal utility} (e.g., quantity). As described in Definition 2, the overall utility of a pattern in the processed database is the cumulative utilities of this pattern in each transaction where it appears in. The \textit{average utility} \cite{hong2011effective} is divide by the length of pattern, and used to avoid the effect of overall utility increasing with the length of pattern. Expected/potential utility determines both uncertainty and utility of a pattern in uncertain data \cite{lin2016efficient2}. Thus, for UPM, this measure is suitable for dealing with uncertain data. Affinitive utility \cite{lin2017fdhup} is proposed to address the special task of correlated UPM, but not used for the general task of UPM. The utility occupancy \cite{gan2019huopm} is more suitable than the \textit{utility} concept and \textit{average utility} for discovering the high-utility patterns which have high utility contribution.

%% file: 3.basicupm.tex
AEI\section{Basic Approaches for High-Utility Itemset Mining}
\label{sec:basicapproaches}

\begin{figure*}[!htbp]
	\setlength{\abovecaptionskip}{0pt}
	\setlength{\belowcaptionskip}{0pt}	
	\centering
	\includegraphics[width=7.0in]{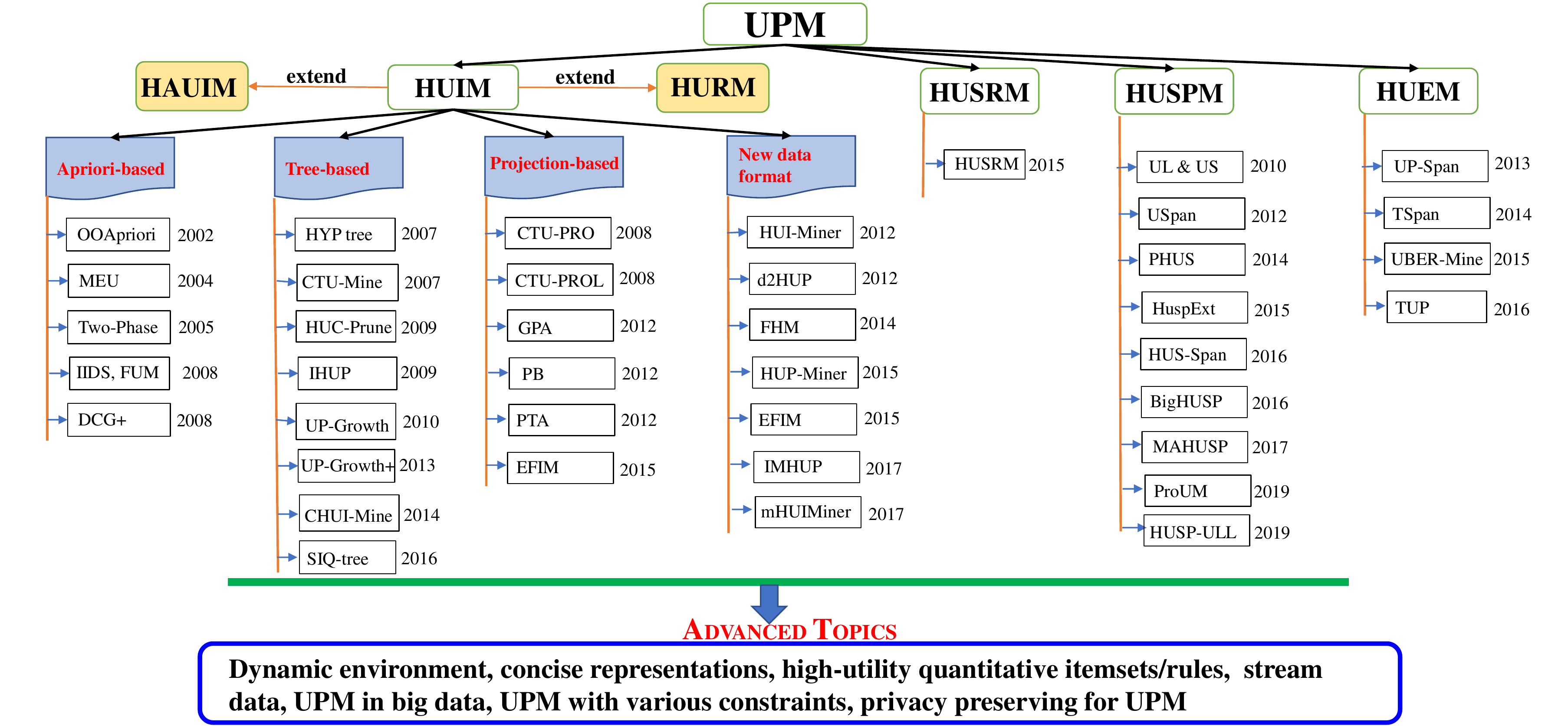}
	\caption{Taxonomy of UPM algorithms.}
	\label{fig:TaxonomyOfUPM}
\end{figure*}

\subsection{Overview of Proposed Categorization}
The development of the utility-oriented algorithms has always been an important issue in data mining area. During the past decades, a significant number of utility-oriented algorithms has been proposed to mine utility-based patterns from various types of data (i.e., transaction data \cite{agrawal1993mining}, sequential data \cite{agrawal1995mining}, episode data \cite{mannila1997discovery}, stream data \cite{golab2003issues,chi2004moment}, etc). Considering that it is infeasible to go through all existing algorithms within a limited space, in this review we select some representative HUIM algorithms. According to the different mining principles and data structures, Fig. \ref{fig:TaxonomyOfUPM} presents a rough overview of techniques to address the UPM problem. Specifically, to facilitate our discussion, we classify these efforts into the following categories: 1) Apriori-based approaches; 2) tree-based approaches; 3) projection-based approaches; and 4) vertical-/horizontal-data-based approaches.

\subsection{Apriori-based Approaches}

In 1994, Agrawal and Srikant proposed the well-known downward closure property, also known as the Apriori property \cite{agrawal1993mining}, which states that all non-empty subsets of a frequent itemset must also be frequent, and any superset of an infrequent itemset cannot be frequent. For example, assuming $\{a, b, c\}$ is frequent, all of its sub-itemsets, such as $\{a, c\}$ and $\{b, c\}$, are also frequent. If $\{d, e\}$ is infrequent, its supersets, such as $\{a, d, e\}$ and $\{d, e, f\}$, are not frequent. Some Apriori-based approaches for HUIM have been further developed. The core step of these Apriori-based UPM algorithms is the generation of candidate $k$-itemsets $C_k$ from high-utility ($k$-1)-itemsets (denoted as $HUI_{k-1}$), and it consists of two operations: join and prune. In join step, the conditional join of two $HUI_{k-1}$ patterns is used to generate candidate set $C_k$. The prune step then reduces the size of $C_k$ by using the utility upper bound (which is similar to the Apriori property \cite{agrawal1993mining}).

$\bullet$   \textbf{OOApriori \& top-$k$ closed utility mining} \cite{shen2002objective,chan2003mining}. In 2002, Shen and Yang proposed an objective-oriented association (OOA) mining approach \cite{shen2002objective}. They integrated the utility constraint into OOApriori (a variant of Apriori \cite{agrawal1993mining}) to prune candidates for deriving the OOA rules. The interestingness of OOA rules are measured in terms of probabilities and utilities in supporting the user's objective. The utility constraint for OOA rules is neither monotone nor anti-monotone. In 2003, Chan \textit{et al}. first defined the concept of utility mining and proposed an objective-directed mining algorithm to mine the top-$k$ closed-utility patterns \cite{chan2003mining}. This was the first time the term ``utility mining" was presented and used to identify both frequent and high-utility itemsets based on business objectives. In this utility-based mining framework, a pruning strategy based on a weak but anti-monotonic condition was developed to reduce search space.

$\bullet$   \textbf{MEU (Mining with Expected Utility)} \cite{yao2004foundational}. In 2005, Yao \textit{et al}. proposed a utility mining model, called mining with expected utility (MEU) \cite{yao2004foundational}, which considers both the purchase quantities (called \textit{internal utility}) and unit profits (called \textit{external utility}) of items to mine HUIs. Note that the term ``mining high-utility itemsets" first appeared in \cite{chan2003mining}, but their concept and definitions were quite different from the definitions of high-utility itemset mining today. It is widely believed that utility-based itemset mining, sequence mining, and web mining originated in \cite{yao2004foundational}. Researchers in the field of UPM consider the MEU model as the first theoretical model and strict definition of high-utility itemset mining. MEU uses a heuristic to determine candidates and usually overestimates. However, it cannot maintain the downward closure property of Apriori \cite{agrawal1993mining}, and the derived results are incomplete.

\begin{table*}[!htbp]
	\centering
	\scriptsize 
	\caption{Apriori-based algorithms for high-utility pattern mining.}
	\label{table_aprioriHUSPM}
	\newcommand{\tl}[1]{\multicolumn{1}{l}{#1}} 
	\begin{tabular}{|l|l|l|l|l|l|} 
		\hline
		\multicolumn{1}{|c|}{\textbf{Name}} & \multicolumn{1}{c|}{\textbf{Description}} & \multicolumn{1}{c|}{\textbf{Pros.}} & \multicolumn{1}{c|}{\textbf{Cons.}} & \multicolumn{1}{c|}{\textbf{Year}} \\ \hline
		
		MEU  \cite{yao2004foundational} & 	\multirow{2}{4cm}{The first theoretical model and strict definitions of high-utility itemset mining.}  & \multirow{2}{4.5cm}{  MEU uses a heuristic to determine candidates and usually overestimates.} & 	\multirow{2}{5cm}{  It cannot maintain the downward closure property of Apriori \cite{agrawal1993mining}, and the derived results are incomplete.}  & 2004  \\ 
		& &  & &   \\   
		& &  & &   \\ \hline

		\multirow{2}{2cm}{UMining \cite{yao2006mining} \& \\  UMining$\_$H \cite{yao2006mining}}    & 	\multirow{2}{4cm}{The general HUIM model with several mathematical properties of the utility measure.} & \multirow{2}{4.5cm}{UMining uses the utility upper bound and UMining$\_$H utilizes a heuristic pruning strategy.} & 	\multirow{2}{5cm}{It generates a large amount of candidate patterns) and suffers from excessive candidate generations and poor scalability.}  & 2006  \\
		& &  & &   \\   
		& &  & &   \\ \hline

		Two-Phase  \cite{liu2005two}  & 	\multirow{2}{4cm}{The TWDC property was proposed to discover HUIs in two phases.} & \multirow{2}{4.5cm}{It can greatly prune a large amount of candidate patterns.} & 	\multirow{2}{5cm}{It suffers from the problem of candidates level-wisely generation-and-test \cite{agrawal1993mining}, and requires multiple database scans.}  & 2005  \\

		& &  & &   \\  
		& &  & &   \\ \hline
		
		\multirow{2}{2cm}{IIDS \cite{li2008isolated}, FUM \cite{li2008isolated}, DCG+ \cite{li2008isolated}} & 	\multirow{2}{4cm}{By applying IIDS to ShFSM and DCG, two methods - FUM and DCG+ - were implemented.}  & \multirow{2}{4.5cm}{For any existing level-wise utility mining method, it can reduce the number of candidates and improve performance.} & 	\multirow{2}{5cm}{It has the same performance issues as Apriori  \cite{agrawal1993mining}.}  & 2008  \\ 
		
		& &  & &   \\  
		& &  & &   \\ \hline		
	\end{tabular}
\end{table*}

$\bullet$   \textbf{UMining and UMining$\_$H} \cite{yao2006mining}. Yao \textit{et al}. then proposed UMining and heuristic UMining$\_$H \cite{yao2006mining} for finding HUIs based on several mathematical properties of the utility measure. The utility constraint is characterized by a property giving the upper bound of the utility value of an itemset. In UMining, the property of utility upper bound  is used as a pruning strategy. UMining$\_$H utilizes another pruning strategy based on a heuristic method \cite{yao2006mining}. However, some HUIs may be erroneously pruned by this heuristic method. Furthermore, neither of them have the downward closure property of Apriori \cite{agrawal1993mining}, and they overestimate too many patterns. Therefore, they suffer from excessive candidate generation and poor scalability.

$\bullet$ \textbf{Two-Phase} \cite{liu2005two}. Note that the downward closure property (w.r.t. the Apriori property \cite{agrawal1993mining}) of the support measure does not hold for the utility. To address the challenge that the utility measure is neither monotone nor anti-monotone, Liu \textit{et al}. proposed the well-known Two-Phase algorithm \cite{liu2005two}. Two-Phase introduces a novel concept named the transaction-weighted downward closure (TWDC) property (for any itemset $X$, if $X$ is not a HTWUI, any superset of $X$ is not an HUI) and used it to discover HUIs in two phases. Phase 1: it finds each itemset $X$ such that \textit{TWU}$(X) \geq minutil$ using the \textit{TWU} upper bound to prune the search space. Initially, it scans a database once to get all 1-itemset $HTWUI_1$; then generates ($k$+1)-level candidate itemsets (with length $k$+1) from length-$k$ candidates $HTWUI_k$ (where $k > 0$). For each iteration, it needs to examine the \textit{TWU} values of candidates by scanning the database once. Finally, it is terminated when no candidate can be generated. Phase 2: it scans the database again to calculate the exact utility of each candidate in the set of $HTWUI_k$ and then outputs the desired HUIs.

$\bullet$   \textbf{IIDS, FUM and DCG+} \cite{li2008isolated}. The itemset share mining problem \cite{li2005direct} can be converted to the utility mining problem by replacing the frequency value of each item in every transaction by its total profit (i.e., multiplying the frequency value by its unit profit). The isolated items discarding strategy (IIDS) \cite{li2008isolated} reduces the number of candidates in every database scan.  By discarding isolated items to reduce the number of candidates and to shrink the database scan in each pass, IIDS can improve the level-wise, multi-pass candidate-generation process. Applying IIDS to ShFSM \cite{li2005direct}  and DCG \cite{li2005direct}, Fast Utility Mining (FUM) \cite{li2008isolated} and Direct Candidates Generation (DCG)+ \cite{li2008isolated} were further developed. The results showed that FUM and DCG+ \cite{li2005direct} are better than MEU, UMining, UMining$\_$H and Two-Phase. However, both still suffer from  the problem of generating and testing candidates in a level-wise way and require multiple database scans.

\textbf{Discussions}. In summary, all of the early UPM approaches improved on the Apriori work \cite{agrawal1993mining}. As shown in Table \ref{table_aprioriHUSPM}, an important drawback is that all of them need to generate a huge amount of candidates since they rely on a loose upper bound on the utilities of candidates. As a result, these approaches may suffer from long execution times (computationally expensive) and consume a huge amount of memory. Moreover, all these algorithms suffer from the same limitations as Apriori-based ARM algorithms, which are to generate candidates not appearing in the database and to perform multiple database scans to mine the desired information. The computational complexity of these Apriori-based UPM techniques depends on the level-wise manner that generates a huge number of candidates. These techniques may have quadratic complexity if the processed data containing long transactions or a low \textit{minutil} threshold value is used.

\subsection{Tree-Based Pattern-Growth Approaches} 
Many early HUIM approaches perform a level-wise exploration of the search space to find HUIs. To avoid the drawback of an Apriori-based level-wise search, several tree-based HUIM algorithms were then proposed to efficiently mine HUIs based on the TWDC property and the pattern-growth-mining approach. Generally speaking, the tree-based UPM algorithms are inspired by the traditional tree-based FPM algorithms, i.e., FP-Growth \cite{han2004mining}. Some mathematical formalism of the pattern-growth method with FP-tree can be referred to \cite{han2004mining}.

$\bullet$   \textbf{HYP tree \cite{hu2007high} and HUC-Prune \cite{ahmed2009efficient3}}. In 2007, Hu \textit{et al}. proposed an approximation method that identifies the contribution of the predefined utility, objective function, and performance metric, and can take advantage of item attributes \cite{hu2007high}. It identifies high-utility combinations and then finds HUIs through a high-yield partition (HYP) tree \cite{hu2007high}. In contrast to the traditional FPM and ARM techniques, its goal is to find segments of data and combinations of items/rules that satisfy certain conditions and maximize a predefined objective function. Different from the former UPM approaches, it conducts ``rule-discovery" with respect to individual attributes and the overall criterion for the discovered results. It aims at mining groups of patterns that when combined, contribute the most to an objective function \cite{hu2007high}. Since the Apriori-like HUIM algorithms suffer from the candidate generation-and-test problem, Ahmed \textit{et al}. proposed a novel tree-based algorithm named High-Utility Candidate Prune (HUC-Prune). The proposed HUC-tree is a prefix tree storing the candidate items in descending order of \textit{TWU} values. Each node in the HUC-tree consists of the item's name and its \textit{TWU} value. Similar to IHUP \cite{ahmed2009efficient}, HUC-Prune replaces the level-wise candidate generation process by a pattern-growth mining approach. It needs at most three database scans for mining the HUIs, and has better performance than the Apriori-based algorithms.

\begin{figure}[!htbp]
	\setlength{\abovecaptionskip}{0pt}
	\setlength{\belowcaptionskip}{0pt}	
	\centering
	\includegraphics[width=3.1in]{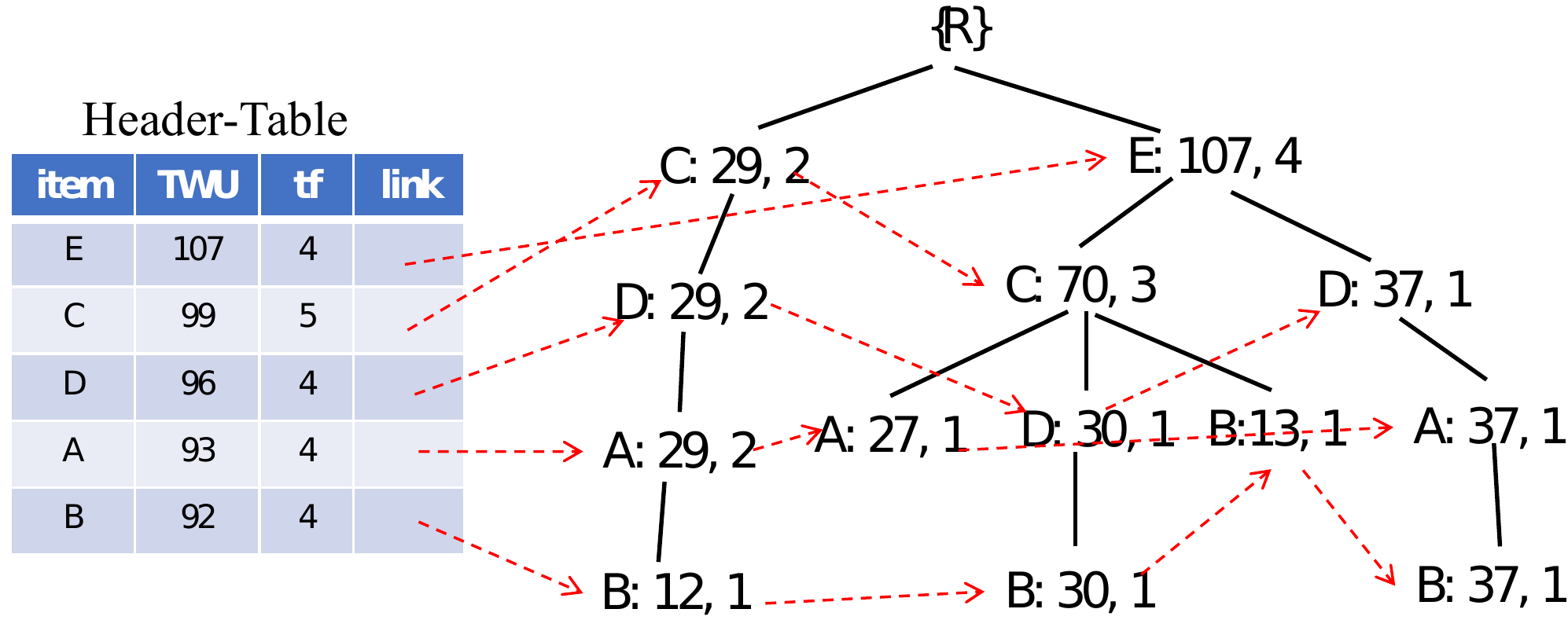}
	\caption{Example IHUP$_{\rm{TWU}}$-tree structure (used by IHUP \cite{tseng2013efficient}).}
	\label{fig:example_IHUPtree}
\end{figure}

$\bullet$   \textbf{IHUP (Incremental High-Utility Pattern mining)} \cite{ahmed2009efficient}. The tree-based algorithm named IHUP with three tree structures (IHUP$_{\rm{PL}}$-tree, IHUP$_{\rm{TF}}$-tree and IHUP$_{\rm{TWU}}$-tree) was proposed for incremental and interactive high-utility pattern mining \cite{ahmed2009efficient}. Each node in the IHUP-tree represents an itemset and consists of the itemset's name, a TWU value, and a support count. The IHUP algorithm has three steps: 1) construction of IHUP-tree, 2) generation of HTWUIs \cite{liu2005two}, and 3) identification of high-utility itemsets. Step 1 is similar to the construction of FP-tree \cite{han2004mining} and a complete illustrated example had been given in \cite{ahmed2009efficient,gan2018survey}. In each transaction, the set of $HTWUI_1$  are sorting in \textit{TWU}-descending order and then continuously inserted into the prefix-based IHUP-tree. Fig. \ref{fig:example_IHUPtree} shows the constructed IHUP-tree for the example database given in \cite{tseng2013efficient}. In step 2, all $HTWUI_k$ are generated from the constructed IHUP-tree using the FP-Growth \cite{han2004mining} approach.  In step 3, all HUIs and their utilities can be identified from the set of HTWUIs by scanning the database once. Thus, IHUP can avoid generating candidates in a level-wise way. Although IHUP significantly outperforms IIDS \cite{li2008isolated}  and Two-Phase \cite{liu2005two}, it still produces too many HTWUIs in step 1. Note that both IHUP and Two-Phase use the \textit{TWU} concept to overestimate the utilities of itemsets. Thus, they produce the same huge number of HTWUIs in step 1. Such a large number of HTWUIs substantially degrades the mining performance in terms of execution time and memory consumption. Moreover, the performance of step 2 is affected by the number of HTWUIs. The reason is that the more HTWUIs is generated in step 1, the longer execution time required for mining HUIs in step 2.

\begin{figure}[!htbp]
	\setlength{\abovecaptionskip}{0pt}
	\setlength{\belowcaptionskip}{0pt}	
	\centering
	\includegraphics[width=3.1in]{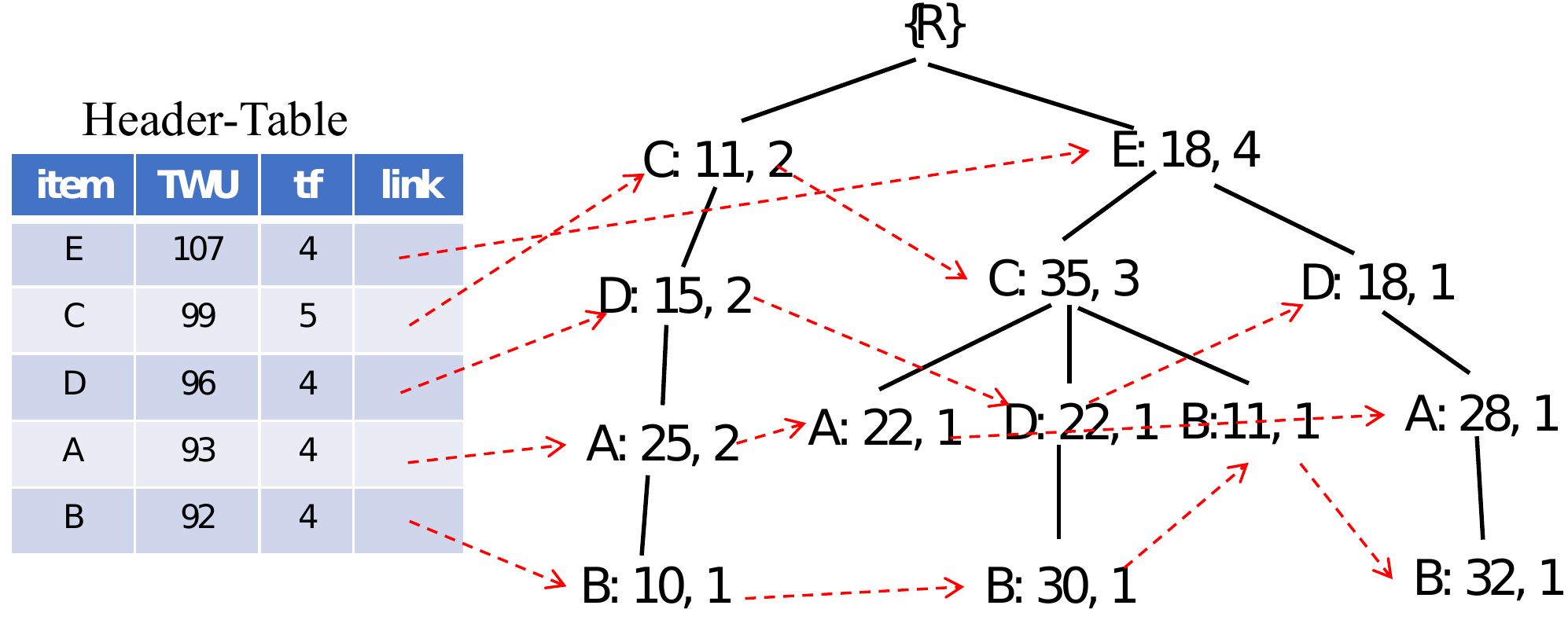}
	\caption{Example UP-tree structure (used by UP-Growth algorithm \cite{tseng2013efficient}).}
	\label{fig:example_UPtree}
\end{figure}

$\bullet$   \textbf{UP-Growth \cite{tseng2010up} and UP-Growth+ \cite{tseng2013efficient}}. Tseng \textit{et al}. designed a more compressed utility-pattern tree (UP-tree) and proposed the well-known utility pattern-growth algorithm (UP-Growth) \cite{tseng2010up} to efficiently mine HUIs. UP-Growth is inspired by the frequency-based FP-Growth method. It integrates four novel strategies, named DLU (Discarding Local Unpromising items), DLN (Decreasing Local Node utilities), DGU (Discarding Global Unpromising items during the construction of a global UP-tree), and DGN (Decreasing Global Node utilities during the construction of a global UP-tree). After two scans of the original database, the UP-tree can be constructed. In the first scan, the utility of each transaction and TWU of each single item are calculated. Discarding global unpromising items, those unpromising items that are not HTWUIs are removed from each transaction, and utilities are eliminated again. Then, the remaining promising items in each transaction are sorted in the descending order of TWU. In the second scan, transactions are inserted into UP-tree by using DGU and DGN strategies. After building the complete UP-tree, as shown in Fig. \ref{fig:example_UPtree}, the potential HUIs (PHUIs) are generated from the global UP-tree with DLU and DLN strategies. In summary, the framework of UP-Growth consists of three steps: 1) scan the database twice to construct a global UP-tree with the DLU and DLN strategies; 2) recursively generate PHUIs from global UP-tree and local UP-trees by UP-Growth with the DGU and DGN strategies; and 3) identify final HUIs from the set of PHUIs. As an improvement of UP-Growth, UP-Growth+ \cite{tseng2013efficient} was then developed by utilizing the minimal utilities of each node in each path in the UP-tree. Compared with UP-Growth, the enhanced UP-Growth+ can decrease the overestimated utilities of PHUIs and greatly reduce the number of candidates.

$\bullet$  \textbf{CHUI-Mine} \cite{song2014mining}. Later, Song \textit{et al}. proposed the CHUI-Mine (Concurrent High-Utility Itemset Mine) algorithm for mining HUIs by dynamically pruning the CHUI-tree structure. The CHUI-tree is introduced to capture the important utility information of the candidate itemsets. By recording changes in support counts of candidates during the tree construction process, it uses dynamic CHUI-tree pruning. CHUI-Mine makes use of a concurrent strategy, enabling the simultaneous construction of a CHUI-tree and the process for discovering HUIs. Thus, it can reduce the problem of huge memory usage for tree construction and traversal in tree-based HUIM algorithms. Concurrent processes generally interact through the following two mechanisms: shared variables and message passing \cite{song2014mining}. Extensive experimental results show that CHUI-Mine is both more efficient and more scalable than Two-Phase \cite{liu2005two}, FUM \cite{li2008isolated}, and HUC-Prune \cite{ahmed2009efficient3}. However, the faster tree-based algorithms, including IHUP \cite{ahmed2009efficient}, UP-Growth \cite{tseng2010up}, and UP-Growth+ \cite{tseng2013efficient} were not all compared.

$\bullet$  \textbf{SIQ-tree (Sum of Item Quantities-tree)} \cite{ryang2016fast}. Tree construction with a single database scan is significant since a database scan is a time-consuming task. In utility mining, an additional database scan is necessary to identify actual high-utility patterns from candidates. A novel tree structure, namely SIQ-tree (Sum of Item Quantities tree) \cite{ryang2016fast}, was developed to capture database information through a single pass. Moreover, a restructuring method is proposed with strategies for reducing overestimated utilities. It can construct the SIQ-tree with only a single scan and decrease the number of candidate patterns effectively with the reduced overestimation utilities, through which mining performance is improved. This approach is faster than IHUP \cite{ahmed2009efficient} on most datasets, but the faster UP-Growth \cite{tseng2010up} and UP-Growth+ \cite{tseng2013efficient}, were not compared.

\begin{table*}[!htbp]
	\centering
	\scriptsize
	\caption{Tree-based pattern-growth algorithms for high-utility pattern mining.}
	\label{table_treeHUSPM}
	\newcommand{\tl}[1]{\multicolumn{1}{l}{#1}} 
	\begin{tabular}{|c|l|l|l|l|l|} 
	\hline
	\multicolumn{1}{|c|}{\textbf{Name}} & \multicolumn{1}{c|}{\textbf{Description}} & \multicolumn{1}{c|}{\textbf{Pros.}} & \multicolumn{1}{c|}{\textbf{Cons.}} & \multicolumn{1}{c|}{\textbf{Year}} \\ \hline

		HYP tree \cite{hu2007high} & 	\multirow{2}{4.5cm}{An approximation method identifies the utility contribution.}  & \multirow{2}{4.5cm}{It can find segments of data through combinations of few items/rules.} & 	\multirow{2}{4cm}{The built HYP tree is huge, and the  memory is costly.}  & 2007  \\ 
		& &  & &   \\ \hline
		
		CTU-Mine \cite{erwin2007ctu}  &  \multirow{2}{4.5cm}{A pattern-growth approach based on a compact data representation named CTU-tree for utility mining.}  & \multirow{2}{4.5cm}{The pattern growth, which avoids candidate generation-and-test, is suitable for dense data.} & 	\multirow{2}{4cm}{The CTU-tree is complex and stores too much information, which may consume much memory.}  & 2007  \\
		& &  & &   \\   
		& &  & &   \\ \hline

		HUC-Prune  \cite{ahmed2009efficient3}  & 	\multirow{2}{4.5cm}{High-utility candidate prune without level-wise candidate generation-and-test.} & \multirow{2}{4.5cm}{It replaces the level-wise candidate generation process by a pattern-growth mining approach.} & 	\multirow{2}{4cm}{The upper bound is high, and the constructed tree is huge.}  & 2009  \\
		& &  & &   \\    
		& &  & &   \\ \hline

		IHUP \cite{ahmed2009efficient} & 	\multirow{2}{4.5cm}{A tree-based approach for incremental and interactive high-utility pattern mining.}  & \multirow{2}{4.5cm}{The IHUP-tree is more compact than previous trees, and IHUP is significantly faster than IIDS and Two-Phase.} & 	\multirow{2}{4cm}{It still uses the TWU and produces too many HTWUIs in phase 1.}  & 2009  \\ 
		& &  & &   \\   
		& &  & &   \\ \hline

		UP-Growth \cite{tseng2010up}  & 	\multirow{2}{4.5cm}{The utility pattern growth algorithm with more compressed utility pattern tree (UP-tree).}  & \multirow{2}{4.5cm}{UP-tree is more compact than IHUP-tree, and the strategies are powerful to reduce the	number of candidates.} & 	\multirow{2}{4cm}{It is time-consuming to recursively process all conditional prefix trees for candidate generation.}  & 2010  \\ 
		& &  & &   \\  
		& &  & &   \\  
		& &  & &   \\ \hline

		UP-Growth+ \cite{tseng2013efficient} & 	\multirow{2}{4.5cm}{An improved version of UP-Growth with two pruning strategies.}  & \multirow{2}{4.5cm}{The enhanced UP-Growth+ can decrease the overestimated utilities of PHUIs and greatly reduce the number of candidates.} & 	\multirow{2}{4cm}{It is time-consuming to recursively process all conditional prefix trees for candidate generation.}  & 2013  \\ 
		& &  & &   \\   
		& &  & &   \\ \hline
		
		CHUI-Mine  \cite{song2014mining} & 	\multirow{2}{4.5cm}{Concurrent High-Utility Itemset Mine (CHUI-Mine) algorithm by dynamically pruning the CHUI-tree.}  & \multirow{2}{4.5cm}{Two mechanisms, shared variables and message passing, can make the CHUI-tree more compact.} & 	\multirow{2}{4cm}{Recursively processing all conditional prefix trees is time-consuming.}  & 2014  \\ 
		& &  & &   \\  
		& &  & &   \\ \hline
		
		SIQ-tree \cite{ryang2016fast} & \multirow{2}{4.5cm}{Sum of Item Quantities.} &  \multirow{2}{4.5cm}{Constructs the SIQ-tree with only a single scan and decreases the number of candidate patterns.} & 	\multirow{2}{4cm}{Recursively processing all conditional prefix trees is time-consuming.}  & 2016  \\ 
		& &  & &   \\   
		& &  & &   \\ \hline
	\end{tabular}
\end{table*}

\textbf{Discussions}.  Characteristics and differences of these tree structures are presented in Table \ref{table_treeHUSPM}. In addition, there are various other utility-based mining methods based on tree structures \cite{erwin2007ctu,lin2011effective}. These tree-based algorithms comprise three steps: 1) construction of trees, 2) generation of candidate HUIs from the trees using the designed pattern-growth approach, and 3) identification of HUIs from the set of candidates. Although these trees are compact, they may not be minimal and still occupy a large memory space. The mining performance is closely related to the number of conditional trees constructed during the entire mining process and the construction/traversal cost of each conditional tree. When using these algorithms on a large database with a low-utility threshold, the storage and traversal costs of numerous conditional trees are high. Thus, one of the performance bottlenecks of these algorithms is the generation of a huge number of conditional trees, which has high time and space costs. 

In summary, to address the disadvantages of the Apriori-like UPM algorithm, and to improve efficiency, the advantages of pattern-growth tree-based techniques are as follows: 1) only need two or three passes over dataset; 2) ``compresses" datasets into the tree structure; 3) no candidate generation; and 4) much faster than Apriori-like approaches. However, they still have some disadvantages: 1) the constructed tree may not fit in memory; 2) the designed tree is expensive to build; 3) it is time-consuming to recursively process all conditional prefix trees to generate candidates; and 4) the constructed tree is sensitive to the \textit{minutil} parameter.

\subsection{Projection-Based Pattern-Growth Approaches} 
In the past, some of projection-based techniques have been commonly used in data mining, i.e., FP-Growth \cite{han2004mining} for FPM and PrefixSpan \cite{han2001prefixspan} for SPM. The general idea of projection-pattern mining is to use target items to recursively project the processed database into some smaller projected sub-databases, and then grow the itemset or subsequence fragments in each projected sub-database. To overcome the disadvantages of the tree-based HUIM approaches, some projection-based techniques have been developed for UPM. Some basic mathematical formalism of projection and projected sub-databases are skipped here, and details can be referred to \cite{han2001prefixspan,lan2012thesis}.

$\bullet$ \textbf{CTU-PRO and CTU-PROL} \cite{erwin2008efficient}. In 2007, Erwin \textit{et al}. first proposed a projection-based CTU-PRO algorithm for HUIM. It mines HUIs by bottom-up traversal of a compressed utility pattern tree (CUP-tree) \cite{erwin2008efficient}, which is a variant of CTU-tree \cite{erwin2007ctu}. The mining of a subdivision from CUP-tree consists of three steps: 1) Construction of ProItem-Table, 2) construction of ProCUP-tree, and 3) mining by ProCup-tree traversal. CTU-PRO creates a GlobalCUP-tree from the transaction database after identifying the individual HTWUIs \cite{liu2005two} with the concept of TWU. For each HTWUI, a smaller projection tree called the LocalCUP-tree is extracted from the GlobalCUP-tree for returning all high-utility itemsets with that item as prefix.  CTU-PRO constructs parallel sub-divisions on disk that can be mined independently. The performance of CTU-PRO is better than Two-Phase \cite{liu2005two} and CTU-Mine \cite{erwin2007ctu}. CTU-PROL introduces two new concepts, compressed transaction utility-prol and CUP-tree, which are used for parallel projection of the transaction database. Note that the anti-monotone property of TWU is used to prune the search space of sub-divisions in CTU-PROL. However, unlike Two-Phase, it avoids a rescan of the database to calculate the actual utilities of HTWUIs. The results show that CTU-PROL outperforms Two-Phase \cite{liu2005two} and CTU-Mine \cite{erwin2007ctu}.

\begin{table*}[!htbp]
	\centering
	\scriptsize
	\caption{Projection-based pattern-growth approaches for UPM.}
	\label{table_projectHUSPM}
	\newcommand{\tl}[1]{\multicolumn{1}{l}{#1}} 
	\begin{tabular}{|c|l|l|l|l|l|} 
		\hline
		\multicolumn{1}{|c|}{\textbf{Name}} & \multicolumn{1}{c|}{\textbf{Description}} & \multicolumn{1}{c|}{\textbf{Pros.}} & \multicolumn{1}{c|}{\textbf{Cons.}} & \multicolumn{1}{c|}{\textbf{Year}} \\ \hline
		
		\multirow{2}{2cm}{CTU-PRO \cite{erwin2008efficient} \& CTU-PROL \cite{erwin2008efficient}}  & 	\multirow{2}{5cm}{Two projection-based algorithms with Compressed Utility Pattern-tree (CUP-tree).}  & \multirow{2}{4.5cm}{They construct parallel sub-divisions on disk that can be mined independently and have good performance.} & 	\multirow{2}{4cm}{TWU is adopted as the upper bound, which generates many redundant candidates.}  & 2008  \\ 
		& &  & &   \\ 
		& &  & &   \\ \hline
		
		GPA and PB \cite{lan2012thesis} &  \multirow{2}{5cm}{Two projection-based mining approaches, GPA (Gradual Pruning Approach) and PB (Projection-Based mining approach).}  & \multirow{2}{4.5cm}{Using projection, they can speed up the runtime and reduce database size when deriving larger itemsets.} & 	\multirow{2}{4cm}{TWU is adopted as the upper bound, which generates many redundant candidates.}  & 2012  \\
		& &  & &   \\     
		& &  & &   \\ \hline
		
		PTA  \cite{lan2012thesis}  & 	\multirow{2}{5cm}{A projection-based upper-bound tightening approach.} & \multirow{2}{4.5cm}{Two effective strategies, named pruning and filtering, are proposed to tighten the upper bounds of utility values.} & 	\multirow{2}{4cm}{The projection of sub-databases is sometimes time-consuming.}  & 2012  \\
		& &  & &   \\    
		& &  & &   \\ \hline
		
	\end{tabular}
\end{table*}

$\bullet$ \textbf{GPA and PB} \cite{lan2012thesis}. Since the tree-based pattern-growth approaches recursively perform tree traversal and generate a series of sub-tree structures, Lan \textit{et al}. proposed two alternative efficient projection-based utility mining approaches, named GPA (Gradual Pruning Approach) \cite{lan2012thesis} and PB (Projection-Based mining approach) \cite{lan2012thesis}. Compared with the level-wise techniques, the property of a projection-based technique is more suitable for improving the utility upper bound. The general idea is to use the overestimated HTWUIs \cite{liu2005two} to recursively project item/sequence databases into some smaller projected databases and grow item/subsequence fragments in each projected sub-database. In addition, PB applies a novel pruning strategy and an indexing mechanism to speed up the runtime and reduce the memory requirement of the mining process. The indexing mechanism imitates traditional projection algorithms (i.e., PrefixSpan \cite{han2001prefixspan}) by projecting sub-databases. Using projection, GPA and PB can significantly reduce database size when deriving larger itemsets and outperform Two-Phase \cite{liu2005two}.

$\bullet$  \textbf{PTA} \cite{lan2012thesis}. Different from PB \cite{lan2012thesis} and GPA \cite{lan2012thesis}, \textit{pruning} and \textit{filtering} strategies are proposed to tighten the upper bounds of utility values in the projection-based upper-bound tightening approach (abbreviated as PTA). The framework of PTA includes the following: 1) finds HTWUIs and high-utility 1-itemsets; 2) performs the pruning strategy and the indexing strategy; 3) projects transactions required by the prefix itemsets to be processed; and 4) finds $k$-HTWUIs and high-utility $k$-itemsets. An effective index mechanism is applied to reduce the time cost of searching relevant transactions that need to be projected in sub-databases. Thus, PTA only needs one database scan. Through experiments, the results show that PTA outperforms the other existing algorithms (i.e., Two-Phase \cite{liu2005two}, GPA \cite{lan2012thesis}, PB \cite{lan2012thesis}, CTU-PRO \cite{erwin2008efficient}, IHUP$_{\rm{PL}}$ \cite{ahmed2009efficient}, IHUP$_{\rm{TWU}}$ \cite{ahmed2009efficient}, and IHUP$_{\rm{TF}}$ \cite{ahmed2009efficient}) in terms of pruning unpromising itemsets, memory usage, and runtime, respectively.

\textbf{Discussions}. In summary, the above UPM approaches, which utilize the database projection mechanism, have the following advantages: 1) mine the complete set of high-utility patterns but reduce the effort of candidate generation; 2) prefix-projection reduces the size of the projected sub-database and leads to efficient processing; and 3) bi-level projection and pseudo-projection may improve mining efficiency, as summarized in Table \ref{table_projectHUSPM}.

\subsection{New Data-Format-Based Approach}
To achieve more efficiency than the tree-based UPM approaches, some algorithms that mine high-utility itemsets using a vertical or horizontal data structure with a single phase were proposed recently, such as HUI-Miner \cite{liu2012mining}, FHM \cite{fournier2014fhm}, d2HUP \cite{liu2012direct}, HUP-Miner \cite{krishnamoorthy2015pruning}, and EFIM \cite{zida2015efim}. Both d2HUP and EFIM use a horizontal database, while others use the vertical data structure. All these algorithms cannot only avoid the disadvantages of Apriori-based approaches but also avoid the disadvantages of the tree-based HUIM approaches. Details are shown in Table \ref{table_listHUPM} and described below.

\begin{table*}[!htbp]
	\centering
	\scriptsize
	\caption{Utility-list-based algorithms for high-utility pattern mining.}
	\label{table_listHUPM}
	\newcommand{\tl}[1]{\multicolumn{1}{l}{#1}} 
	\begin{tabular}{|c|l|l|l|l|l|} 
	\hline
	\multicolumn{1}{|c|}{\textbf{Name}} & \multicolumn{1}{c|}{\textbf{Description}} & \multicolumn{1}{c|}{\textbf{Pros.}} & \multicolumn{1}{c|}{\textbf{Cons.}} & \multicolumn{1}{c|}{\textbf{Year}} \\ \hline
	
		HUI-Miner \cite{liu2012mining} & 	\multirow{2}{4cm}{The first one-phase model to mine high-utility itemsets.}  & \multirow{2}{4.5cm}{It first introduced the concept of the remaining utility and the vertical data structure w.r.t. utility-list.} & 	\multirow{2}{4.5cm}{The join operations between utility lists of ($k$+1)-itemsets and $k$-itemsets is time-consuming.}  & 2012  \\ 
		& &  & &   \\   
		& &  & &   \\ \hline

		d2HUP \cite{liu2012direct}  & 	\multirow{2}{4cm}{Another algorithm that can directly discover HUIs without maintaining candidates.} & \multirow{2}{4.5cm}{It efficiently obtains the utility of each enumerated itemset and the upper bound on utilities using the CAUL.} & 	\multirow{2}{4.5cm}{The tree structure and CAUL consume more memory.}  & 2012  \\
		& &  & &   \\    
		& &  & &   \\ \hline

		FHM \cite{fournier2014fhm} &  \multirow{2}{4.5cm}{An improved version of HUI-Miner with a pruning strategy named EUCP.}  & \multirow{2}{4.5cm}{It not only has the advantages of HUI-Miner but also reduces the join operations between utility lists.} & 	\multirow{2}{4.5cm}{It consumes slightly more memory than HUI-Miner and has poor performance on dense datasets.}  & 2014  \\
		& &  & &   \\    
		& &  & &   \\ \hline

		HUP-Miner \cite{krishnamoorthy2015pruning} & \multirow{2}{4cm}{An improved version of HUI-Miner with two new pruning strategies (PU-Prune and LA-Prune).} &  \multirow{2}{4.5cm}{The two new pruning strategies can reduce the join operations between utility lists.} & 	\multirow{2}{4.5cm}{It needs to explicitly set the number of dataset partitions, while these partitions cannot always improve the efficiency.}  & 2015  \\ 
		& &  & &   \\   
		& &  & &   \\ \hline

		EFIM \cite{zida2015efim} & 	\multirow{2}{4cm}{Uses projection and transaction-merging techniques for reducing the cost of database scans.}  & \multirow{2}{4.5cm}{It consumes less memory, and its complexity is roughly linear with the number of items in the search space.} & 	\multirow{2}{4.5cm}{Sometimes the recursive projection is time-consuming and uses a lot of memory.}  & 2015  \\ 
		& &  & &   \\  
		& &  & &   \\ \hline

		IMHUP  \cite{ryang2017indexed} & 	\multirow{2}{4cm}{A novel utility-list-based algorithm for HUPs mining without any candidate generation.}  & \multirow{2}{4.5cm}{It uses the indexed utility list to reduce the join operations between utility lists.} & 	\multirow{2}{4.5cm}{The upper bound on utilities is not tight enough.}  & 2017  \\ 
		& &  & &   \\   
		& &  & &   \\ \hline
				
		mHUIMiner \cite{peng2017mhuiminer}  & 	\multirow{2}{4cm}{An efficient one-phase algorithm that combines some ideas from HUI-Miner and IHUP.}  & \multirow{2}{4.5cm}{It utilizes utility list and remaining utility and performs well on sparse datasets.} & 	\multirow{2}{4.5cm}{It still suffers from some problems similar to those of HUI-Miner and IHUP.}  & 2017  \\ 
		& &  & &   \\  
		& &  & &   \\ \hline

	\end{tabular}
\end{table*}

$\bullet$  \textbf{HUI-Miner  \cite{liu2012mining} and FHM  \cite{fournier2014fhm}}. HUI-Miner (High-Utility Itemset Miner) \cite{liu2012mining} is the first one-phase algorithm to discover HUIs. It proposes a vertical data structure named utility-list \cite{liu2012mining} and the concept of \textit{remaining utility} \cite{liu2012mining}, which have been widely extended in many other newly UPM algorithms. As a compact data structure, utility-list can store utility information for the potential patterns that may have high utility value. The utility-list of an itemset $X$ in a database $D$ is a set of tuples corresponding to the transactions in which $X$ appears. Each tuple is defined as $<$$tid, iu, nu$$>$ for every transaction $T_q$ containing $X$, in which the \textit{tid} element is the transaction identifier of $T_q$, the \textit{iu} element is the utility value of $X$ in $T_q$, and \textit{ru} element is the \textit{remaining utility} value of $X$ in $T_q$. More details about the \textit{remaining utility}, utility-list structure, and its construction can be referred to \cite{liu2012mining}. The construction process of utility-list is quite efficient and consumes little memory.  By keeping necessary information from the transaction database in memory, HUI-Miner can directly mine HUIs by spanning the search space w.r.t. a set-enumeration tree \cite{rymon1992search}.  As an enhanced version of HUI-Miner \cite{liu2012mining}, FHM (Fast High-Utility Miner) \cite{fournier2014fhm} utilizes a novel pruning strategy named EUCP (Estimated Utility Co-occurrence Pruning) to reduce the costly join operations of utility-lists. EUCP is based on the Estimated Utility Co-Occurrence Structure (EUCS) \cite{fournier2014fhm}. Using utility-list \cite{liu2012mining}, HUI-Miner and FHM need only two database scans to construct a series of utility-lists of $HTWUI_1$. Then, utility-lists of ($k$+1)-itemsets can be obtained by performing the join operations of utility-lists of $k$-itemsets.  They can directly discover HUIs by keeping utility-list in memory, and utilizes the upper bound of the remaining utility.  HUI-Miner and FHM outperform than the all previous algorithms on most datasets, in terms of running time (almost two orders of magnitude faster) and memory cost. FHM is more faster than HUI-Miner \cite{liu2012mining}, especially for dense databases, but not efficient for databases that are sparse. However, the drawback is that both of them need to perform costly join operations among a series of utility-lists, which can be time costly. Note that some quantitative results are already reported on the same benchmark datasets in \cite{liu2012mining,fournier2014fhm}.

$\bullet$   \textbf{d2HUP} \cite{liu2012direct}. d2HUP is also able to directly discover HUIs without candidate generation. It utilizes another novel data structure, named CAUL (Chain of Accurate Utility Lists) \cite{liu2012direct} to store the necessary information. In contrast to HUI-Miner, it enumerates an itemset as a prefix extension of its prefix itemset. In fact, the search space of d2HUP is a variant of set-enumeration tree \cite{rymon1992search}. It can efficiently calculate the utility of each enumerated itemset and the upper bound on utilities of the prefix-extended itemsets. In fact, d2HUP also utilizes the similar concept of \textit{remaining utility} to tighten the utility upper bound, which is much tighter than TWU. This upper bound is tightened by iteratively filtering out irrelevant items when constructing CAUL.  More specifically, it requires less memory than different kinds of tree structures used in the above-mentioned algorithms. d2HUP was shown to be more efficient than Two-Phase \cite{liu2005two}, UP-Growth \cite{tseng2010up}, and HUI-Miner \cite{liu2012mining}, but the performance was not compared with some recent algorithms, such as FHM \cite{fournier2014fhm} and HUP-Miner \cite{krishnamoorthy2015pruning}.

$\bullet$   \textbf{HUP-Miner}. HUP-Miner \cite{krishnamoorthy2015pruning} is an improvement algorithm based on HUI-Miner \cite{liu2012mining}. Two new pruning strategies, PU-Prune (based on dataset partition) and LA-Prune (based on the concept of \textit{lookahead pruning}), are introduced in HUP-Miner to limit the search space for mining HUIs \cite{krishnamoorthy2015pruning}. It needs to set the number of dataset partitions $K$, which determines how many partitions processed internally. However, the optimal value of $K$ is hard to find empirically for a given dataset. Based on the concept of remaining utility \cite{liu2012mining}, LA-Prune provides a tighter utility upper bound of any $k$-itemset. Thus, a huge number of unpromising $k$-itemset ($k$ $\geq$ 2) that have low utility can be pruned. It has been shown that HUP-Miner is significantly faster than HUI-Miner. In fact, the PU-Prune strategy based on dataset partition does not always have an effect on runtime and memory consumption. In addition, a shortcoming is that the number of partitions is required to be set explicitly by users, since it is an additional parameter.

$\bullet$  \textbf{IHUI-Mine (Index High-Utility Itemsets Mine)} \cite{song2016high}. As mentioned before, these candidate generation-and-test approaches suffer from the drawbacks of having an immense candidate pool and requiring several database scans. Meanwhile, methods based on pattern growth tend to consume large amounts of memory to store conditional trees. IHUI-Mine uses the subsume index \cite{song2008index}, a data structure for efficient frequent itemset mining, to enumerate the desired HUIs and prune the search space. The experimental results show that IHUI-Mine outperforms some popular algorithms, including Two-Phase \cite{liu2005two}, FUM \cite{li2008isolated}, and HUC-Prune \cite{ahmed2009efficient3}, but it has not been compared with the state-of-the-art algorithms.

$\bullet$   \textbf{IMHUP} \cite{ryang2017indexed}. In the framework of list-based high-utility pattern mining, there are a number of comparison and join operations of entries within lists causing enormous execution time costs. Based on the indexed utility-list (IU-list) \cite{ryang2017indexed}, two techniques were developed in IMHUP (Indexed list-based Mining of High-Utility Patterns) to reduce utility upper-bounds that satisfy the anti-monotonic property. IMHUP-RUI and IMHUP-CHI \cite{ryang2017indexed} generate high-utility patterns without any construction of additional local-lists when the current lists only contain information of the same revised transactions. They further utilize the upper-bound utilities in IU-lists to decrease the search space.

$\bullet$   \textbf{EFIM} \cite{zida2015efim}. The projection-based EFficient high-utility Itemset Mining (EFIM) algorithm introduces several new ideas, including two new upper bounds named revised \textit{sub-tree utility} and \textit{local utility}, and a array-based utility computing technique. To reduce the cost of database scans, EFIM further proposes the database projection and transaction merging techniques named High-utility Database Projection (HDP) and High-utility Transaction Merging (HTM). As larger itemsets are explored, both projection and merging reduce the size of the database. The main ideas of HDP and HTM are described in \cite{zida2015efim}. The time and space complexity of EFIM is roughly linear with the number of distinct items in the search space. The competitive results show that EFIM is in general 2 to 3 orders of magnitude faster than the state-of-the-art algorithms (UP-Growth+ \cite{tseng2013efficient}, HUI-Miner \cite{liu2012mining}, FHM \cite{fournier2014fhm}, d2HUP \cite{liu2012direct}, and HUP-Miner \cite{krishnamoorthy2015pruning}) on dense datasets and performs quite well on sparse datasets.

$\bullet$  \textbf{mHUIMiner (modified HUI-Miner)}. mHUIMiner \cite{peng2017mhuiminer}  is a hybrid algorithm that combines some ideas from HUI-Miner \cite{liu2012mining} and IHUP-tree \cite{ahmed2009efficient}. It adopts the utility-list and remaining utility. It utilizes a tree structure to guide the itemset expansion process, and thus the itemsets that are nonexistent in the database can be avoided. Unlike current techniques, it does not have a complex pruning strategy that requires expensive computational overhead. It was shown to well perform on sparse datasets, and provide the best runtime on sparse datasets, while having a comparable performance than other state-of-the-art algorithms (e.g., HUI-Miner \cite{liu2012mining}, FHM \cite{fournier2014fhm}, and EFIM \cite{zida2015efim}) on dense datasets.

\textbf{Discussions}. All the algorithms discussed in this subsection utilize the new data structure to store necessary information about each itemset. By spanning the search space w.r.t. a set-enumeration tree \cite{rymon1992search}, they can easily calculate the total utility of an itemset by performing join operations of the built utility-lists. Moreover, an upper bound on the overall utilities of itemsets called the remaining utility is calculated using utility-lists. It can be used to determine if each pattern and its extensions are not high-utility itemsets (to reduce the search space). The upper bound with remaining utility is equivalent to the upper bound proposed in d2HUP \cite{liu2012direct}. Although a pattern-growth approach in d2HUP can avoid considering itemsets not appearing in the database, the used hyper-structure still consumes a considerable amount of memory \cite{liu2012direct}. Some competitive results of these UPM methods have been compared and summarized in recent studies \cite{zida2015efim,peng2017mhuiminer}.

As shown at Table \ref{table_listHUPM}, the HUIM algorithms are based on a combination of vertical or horizontal data formats and typical approaches. These hybrid algorithms combine different techniques to mine high-utility patterns in such a way that the strengths of each technique are utilized to maximize their efficiency. The properties of these one-phase algorithms are as follows:

\begin{enumerate}
	\item \textit{Complete result}: The completeness is guaranteed as the traversal of the search space w.r.t. set-enumeration tree.
	\item \textit{Stable result}: The result is stable as all exact utility information is stored in a vertical or horizontal data structure. Depth-first searching is also used to quickly calculate the utilities.
	\item  \textit{Efficiency}: The algorithm is efficient relative to algorithms that traverse the complete search space. Moreover, the sort order of items in set-enumeration tree affects the mining efficiency, but not the final mining results of patterns.
	\item \textit{Parameter sensitivity}: These algorithms, except for HUP-Miner, only have \textit{minutil} as the parameter, and are sensitive to it.
\end{enumerate}

%% file: 4.advanced.tex
\section{Advanced Topic of UPM}
\label{sec:advancedtopic}

\subsection{Mining High Average Utility Itemsets}

A main challenge in HUIM is that the exponential search space for HUIM is extremely large when the number of distinct items or the size of the database is too large. The other challenge is that existing HUIM methods overlook the fact that longer itemsets result in higher utility values. A large itemset may have an unreasonable estimated profit as opposed to its actual value. Therefore, the concept named high average-utility itemset mining (HAUIM) is proposed \cite{hong2011effective}. HAUIM discovers utility patterns by considering both their utilities and lengths, thus providing a different utility measure than traditional HUIM. HAUIM divides the utility of an itemset by its length (the number of items that the itemset contains). Up to now, some interesting works have been extensively studied, such as Apriori-based algorithms \cite{hong2011effective}, projection-based PAI \cite{lan2012efficiently}, utility-list based HAUI-Miner \cite{lin2016efficient3,lin2017ehaupm}, and other hybrid algorithms with different upper-bound models \cite{yun2017damped,lin2017ehaupm}.

\subsection{HUIM in Dynamic Environments}
In a wide range of applications, the processed data may be commonly dynamic but not static. The dynamic data are more complicated and difficult to handle than the static data. Most algorithms process a static database to mine HUIs. In real-world applications, records/transactions are dynamically changed (i.e., inserted, deleted, and modified) in the original database. Some preliminary studies have been done on this issue for UPM.

 $\bullet$  \textbf{Case 1: \textit{HUIM with record insertion}.} Data mining is an iterative process, and incremental data mining \cite{cheung1996maintenance,hong2001new} provides the ability to continuous analyze and mine the data by using previous data structure and mining results. Up to now, some incremental models have been developed for mining HUIs with record insertion, such as IHUP \cite{ahmed2009efficient}, FUP-HUI-INS \cite{lin2012incremental}, PRE-HUI-INS \cite{lin2014incrementally}, HUI-list-INS \cite{lin2015incremental}, and EIHI \cite{fournier2015efficient}. Among these, the early algorithms, e.g., FUP-HUI-INS and PRE-HUI-INS, utilize the  utility-oriented dynamic maintain strategies that are extended by the original FUP \cite{cheung1996maintenance} and pre-large \cite{hong2001new} concepts. Since FUP-HUI-INS and PRE-HUI-INS algorithms are processed by a Two-Phase model, an additional database rescan is still necessary to find the actual HUIs. Furthermore, computations are required to find the HTWUIs based on the pattern-growth approach. Both HUI-list-INS and EIHI  utilize the utility-list \cite{liu2012mining} and utility property to significantly reduce runtime and memory usage. More complete reviews can be referred to  \cite{gan2018survey}.

$\bullet$   \textbf{Case 2: \textit{HUIM with record deletion}.} In practical situations, record deletion is also an important issue in databases. Cheung \textit{et al}. designed the FUP2 concept \cite{cheung1997general} to discover frequently updated itemsets for record deletion. Hong \textit{et al}. developed the pre-large concept \cite{hong2001new} for handling record deletion to avoid a multiple database scan each time. Two support thresholds are separately set in pre-large \cite{hong2001new}, and thus the original database is not required to be scanned until the number of accumulative deleted transactions achieves the designed safety bound. Since the FUP2 concept \cite{cheung1997general} cannot be directly applied to the HUIM, Lin \textit{et al}. separately designed the FUP-HUI-DEL \cite{lin2015mining} and PRE-HUI-DEL \cite{lin2015efficient} algorithms for handling record deletion to maintain and update the new HUIs based on the Two-Phase model. Recently, an efficient dynamic algorithm named HUI-list-DEL \cite{lin2016fast2} was developed to discover HUIs by maintaining the built utility-list \cite{liu2012mining} structure for record deletion in dynamic databases. The new HUIs can be directly produced without candidate generation or numerous database scans.

$\bullet$   \textbf{Case 3: \textit{HUIM with record modification}.} As one of the three common operations (record insertion, deletion, and modification) in databases, record modification is also commonly seen in real-life situations. For example, some typos or errors may occur when the collected data from periodic transactions is input into a computer using a keyboard.  Thus, some information may become invalid or new information may arise. Lin \textit{et al}. first proposed the FUP-HUP-tree-MOD algorithm \cite{lin2016updating} to address this issue. It is based on the FUP concept \cite{cheung1996maintenance} and shows better performance compared to Two-Phase and some tree-based algorithms in batch mode. In addition, a faster PRE-HUI-MOD algorithm \cite{lin2015fast} extends the pre-large concept \cite{hong2001new} to set the effective upper bound for discovering HTWUIs and HUIs from the dynamic databases.

\subsection{Concise Representations of Utility Patterns}

In the field of FPM, many techniques have been devised to derive compact representations of frequent patterns that eliminate redundancy but have rich information, such as free sets \cite{boulicaut2003free}, non-derivable sets \cite{calders2002mining}, maximal itemsets \cite{gouda2001efficiently}, and closed itemsets \cite{pasquier1999efficient}. These representations significantly reduce the number of extracted frequent patterns, but some lead to loss of information (e.g., maximal itemsets \cite{gouda2001efficiently}). Although the above UPM methods perform well in some cases, their performance may degrade when the minimum utility threshold is low. A large number of HUIs and candidates lead to long execution times and huge memory consumption. When computing resources are limited, this is a serious problem for the mining task. However, a large amount of HUIs is difficult to comprehend and be analyzed by users. Thus, it is often impractical to generate and return the entire set of HUIs. 

$\bullet$   \textbf{Maximal high-utility pattern}. To return representative HUIs to users, some concise representations of HUIs were proposed. Chan \textit{et al}. introduced the concept of a utility frequent closed pattern \cite{chan2003mining}, the definition of which is different from high-utility itemset \cite{yao2004foundational,tseng2013efficient}. Shie \textit{et al}. then proposed a new representation called maximal high-utility itemset in which a HUI is not a subset of any other HUI \cite{shie2012efficient}.  Although maximal HUI reduces the number of extracted HUIs, it is not lossless because the utilities of the subsets of a maximal HUI cannot be known without rescanning the database. Moreover, recovering all HUIs from the set of maximal HUIs is very inefficient since many subsets of a maximal HUI may have low utility.

$\bullet$   \textbf{Closed high-utility pattern}. To provide not only compact but also complete information about high-utility itemsets to users, Tseng \textit{et al}. first addressed the problem of redundancy in high-utility itemset mining \cite{tseng2015efficient}.  A lossless and compact representation named closed high-utility itemset \cite{tseng2015efficient} was introduced. To mine this representation, they proposed three algorithms named AprioriHC (Apriori-based approach for mining High-utility Closed itemsets), AprioriHC-D (AprioriHC algorithm with Discarding unpromising and isolated items), and CHUID (Closed High-Utility itemset Discovery) \cite{tseng2015efficient}.  Fournier-Viger \textit{et al}. then proposed a fast and memory efficient algorithm named EFIM-Closed \cite{fournier2016efim} to discover closed HUIs by extending the EFIM model \cite{zida2015efim}. It proposes three strategies to mine CHUIs efficiently: closure jumping, forward closure checking, and backward closure checking. EFIM-Closed relies on two new upper bounds, named local utility and sub-tree utility, to prune the search space, and it can calculate these upper bounds efficiently. Inspired by utility-list \cite{liu2012mining}, some more efficient one-phase algorithms have been proposed to address this interesting issue, such as CHUI-Miner  \cite{wu2015mining} and CHUM \cite{sahoo2016efficient}.

\subsection{Mining High-Utility Quantitative Itemsets/Rules}

Although extensive studies have been proposed for high-utility itemset mining, a critical limitation of these studies is that they ignore the quantity attribute of items in discovered HUIs. However, such information can be very useful and valuable in many applications. In view of this, the concept of High-Utility Quantitative Itemset mining (abbreviated as HUQI)  \cite{yen2007mining,li2014efficient} has emerged. In the framework of HUQI mining, an item may have different quantities in the database and each item carrying a different quantity is regarded as a quantitative item. HUQI \cite{yen2007mining} and more efficient vertical utility-list-based VHUQI \cite{li2014efficient} were thus developed. An example of such a rule is (\textit{bread}, 3, 4) $ \Rightarrow $ (\textit{milk}, 2, 3), which means that most customers who purchased three or four breads also  purchased two or three milks. We can use this information to package products with quantities that have high utility and estimate the number of items that need to be reserved according to the number of other items.

\subsection{High-Utility Sequential Pattern Mining}

By integrating the utility factor and sequence data, the problem of high-utility sequential pattern mining (HUSPM) was introduced. For handling the utility of web log sequences, two tree structures, called utility-based WAS tree (UWAS-tree) \cite{ahmed2011framework2} and incremental UWAS-tree (IUWAS-tree) \cite{ahmed2011framework2}, were developed to mine web access sequences (WASs). However, a sequence element with multiple items, such as [($a$, 3)($c$, 4)], cannot be supported in these two models. The considered scenarios are rather simple, which limits their applicability for handling complex sequences. Then, some algorithms were proposed to address the HUSPM problem.

$\bullet$   \textbf{UL and US} \cite{ahmed2010novel}. Since both UWAS-tree and IUWAS-tree algorithms cannot deal with sequences containing multiple items in each sequence element (transaction), Ahmed \textit{et al}. designed two algorithms (level-wise Utility-Level (UL)  \cite{ahmed2010novel} and pattern-growth Utility-Span (US) \cite{ahmed2010novel}) to mine HUSPs.  UL and US extend traditional sequential pattern mining (SPM). The utility of a sequential pattern is calculated in two ways. The utilities of sequences having only distinct occurrences are added together, while the highest occurrences are selected from sequences with multiple occurrences and used to calculate the utilities. However, the problem definition in UL and US  \cite{ahmed2010novel} is rather specific. No generic framework for transferring from SPM to high-utility sequence analysis has been proposed.

\begin{table*}[!htbp]
	\centering
	\scriptsize 
	\caption{Algorithms for high-utility sequential pattern mining.}
	\label{table_HUSPM}
	\newcommand{\tl}[1]{\multicolumn{1}{l}{#1}} 
	\begin{tabular}{|c|l|l|l|l|l|} 
		\hline
		\multicolumn{1}{|c|}{\textbf{Name}} & \multicolumn{1}{c|}{\textbf{Description}} & \multicolumn{1}{c|}{\textbf{Pros.}} & \multicolumn{1}{c|}{\textbf{Cons.}} & \multicolumn{1}{c|}{\textbf{Year}} \\ \hline
		
		\multirow{2}{2cm}{UWAS-tree \& IUWAS-tree  \cite{ahmed2011framework2}} & 	\multirow{2}{4.5cm}{The specific algorithms designed for mining utility web	log sequences.}  & \multirow{2}{4.5cm}{The first paper that integrates utility measure into sequential pattern mining.} & 	\multirow{2}{4cm}{The considered scenarios are rather simple, which limits their applicability for complex sequences.}  & 2010  \\ 
		& &  & &   \\ 
		& &  & &   \\ \hline

		UL \& US \cite{ahmed2010novel} &  \multirow{2}{4.5cm}{A level-wise Utility-Level (UL) algorithm and a pattern-growth Utility-Span (US) algorithm.}  & \multirow{2}{4.5cm}{They first extend SPM to HUSPM, and introduce the calculation of utility in a sequence.} & 	\multirow{2}{4cm}{The problem definition is rather specific. No generic framework is proposed.}  & 2010  \\
		& &  & &   \\    
		& &  & &   \\ \hline

		USpan \cite{yin2012uspan} & 	\multirow{2}{4.5cm}{First formalizes the problem of HUSPM, and proposes a generic framework.} & \multirow{2}{4cm}{The width and depth pruning methods substantially reduce the search space in the LQS-tree.} & 	\multirow{2}{4cm}{Data representation w.r.t. utility matrix in USpan is quite complex and memory-costly.}  & 2012  \\
		& &  & &   \\   
		& &  & &   \\ \hline

		PHUS \cite{lan2014applying} & \multirow{2}{4.5cm}{A projection-based algorithm for mining HUSPs with the maximum utility measure called sequence-utility upper-bound (SUUB).} &  \multirow{2}{4.5cm}{The projection mechanism and upper-bound SUUB model can avoid considering too many candidates.} & 	\multirow{2}{4cm}{The upper bound on sequence utility is not tight enough, and consumes longer runtime.}  & 2014  \\ 
		& &  & &   \\   
		& &  & &   \\ \hline

		HuspExt  \cite{alkan2015crom} & 	\multirow{2}{4.5cm}{Based on the upper-bound CRoM, it utilizes a Pruning Before Candidate Generation (PBCG) strategy.}  & \multirow{2}{4.5cm}{PBCG strategy prunes some unpromising sequences for mining HUSPs.} & 	\multirow{2}{4cm}{The CRoM upper-bound on sequence utility produces the incomplete set of HUSPs.}  & 2015  \\ 
		& &  & &   \\    
		& &  & &   \\ \hline

		HUS-Span  \cite{alkan2015crom} & 	\multirow{2}{4.5cm}{Two upper-bounds (PEU and RSU) and utility-chain are proposed in this method.}  & \multirow{2}{4.5cm}{PEU is more tighter than previous upper-bounds, and utility-chain is easily implemented.} & 	\multirow{2}{4cm}{It may consume longer runtime than USpan on some datasets.}  & 2016  \\ 
		& &  & &   \\    
		& &  & &   \\ \hline

		ProUM  \cite{gan2019proum} & 	\multirow{2}{4.5cm}{Using a new upper-bound named SEU and the array-based structure named utility-array.}  & \multirow{2}{4.5cm}{SEU is a correct upper-bound and the projection-based pruning strategies are powerful to prune unpromising sequences.} & 	\multirow{2}{4cm}{The SEU upper bound is still an over-estimated bound; the search space can be further reduced.}  & 2019  \\ 
		& &  & &   \\    
		& &  & &   \\ \hline

		HUSP-ULL  \cite{gan2019fast} & 	\multirow{2}{4.5cm}{Based on the PEU upper-bound, it utilizes the utility-linked (UL)-list structure and two pruning strategies to prune candidates.}  & \multirow{2}{4.5cm}{It outperforms the state-of-the-art algorithms for mining HUSPs, in terms of execution time and memory consumption.} & 	\multirow{2}{4cm}{The proposed UL-list structure is a complicated structure for implementation.}  & 2019  \\ 
& &  & &   \\    
& &  & &   \\ \hline

	\end{tabular}
\end{table*}

$\bullet$   \textbf{USpan} \cite{yin2012uspan}. Yin \textit{et al}. then formalized the problem of HUSPM, and proposed a generic framework and the USpan algorithm to mine high-utility sequences. \cite{yin2012uspan}. A lexicographic quantitative sequence tree (LQS-tree) is  constructed as the search space. Two concatenation mechanisms, \textit{I-Concatenation} and \textit{S-Concatenation}, are used to generate newly concatenated utility-based sequences. Based on the LQS-tree structure, USpan \cite{yin2012uspan} adopts the  sequence-weighted utilization (SWU) measure and the Sequence Weighted Downward Closure (SWDC) property to prune unpromising sequences and to improve the mining performance. However, a shortcoming of USpan is that the data representation w.r.t. the utility matrix is quite complex and memory-costly.

$\bullet$  \textbf{PHUS}  \cite{lan2014applying}. Lan \textit{et al}. then proposed the projection-based high-utility sequential pattern mining (PHUS) algorithm for mining HUSPs with the maximum utility measure and a \textit{sequence-utility upper-bound} (SUUB) model  \cite{lan2014applying}. The algorithm extends PrefixSpan \cite{han2001prefixspan} and uses a projection-based pruning strategy to obtain tight upper bounds on sequence utilities. Thus, it can avoid considering too many candidates, and improves the performance of mining HUSPs using the SUUB model.

$\bullet$   \textbf{HuspExt} \cite{alkan2015crom} and \textbf{HUS-Span} \cite{wang2016efficiently}. Alkan \textit{et al}. \cite{alkan2015crom} designed the high-utility sequential pattern extraction (HuspExt) algorithm with an upper-bound called Cumulate Rest of Match (CRoM). It uses a pruning before candidate generation (PBCG) strategy to prune unpromising sequences for mining HUSPs. However, HuspExt cannot discover the complete HUSPs due to the incorrect upper bound.  In view of the previous upper bounds on sequence utilities not being tight enough, HUS-Span \cite{wang2016efficiently} utilizes two tight utility upper bounds, called prefix extension utility (PEU) and reduced sequence utility (RSU), as well as two companion pruning strategies, to identify high-utility sequential patterns.

$\bullet$   \textbf{ProUM} \cite{gan2019proum}.  Gan \textit{et al.} \cite{gan2019proum} proposed an efficient projection-based utility mining approach  named ProUM to discover high-utility sequences by utilizing the upper bound named sequence extension utility (SEU) and the utility-array structure \cite{gan2019proum}. Different from the upper bound used in USpan, SEU can guarantee the  correctness and completeness of discovered results on sequence data. Besides, ProUM has better performance up to two orders of magnitude in terms of execution time on most sequence datasets than USpan and HUS-Span.

$\bullet$   \textbf{HUSP-ULL} \cite{gan2019fast}. The state-of-the-art HUSP-ULL \cite{gan2019fast} algorithm utilizes a new data structure namely utility-linked (UL)-list and two pruning strategies (called look ahead strategy and irrelevant item pruning strategy) to fast discover HUSPs. According to the extensive experiments \cite{gan2019fast}, it shows that HUSP-ULL is the fastest when comparing to the current HUSPM algorithms.  Some competitive results of these recent state-of-the-art HUSPM methods have been compared and summarized in the studies of ProUM \cite{gan2019proum} and HUSP-ULL \cite{gan2019fast}.

Main characteristics of these HUSPM algorithm are summarized in Table \ref{table_HUSPM}. In addition, Shie \textit{et al}. explored a new problem of mining high-utility mobile sequential patterns (HUMSPs) by integrating mobile data mining with utility mining \cite{shie2011mining,shie2013efficient}. This is the first work that combines mobility patterns with utility factor to find high-utility mobile sequential patterns.

\subsection{High-Utility Episode Mining}

When the sequential data becomes an event sequence, the task of frequent episode mining (FEM) \cite{mannila1997discovery} is introduced. FEM reveals a significant amount of useful information hidden in the event sequence with a wide range of applications \cite{mannila1997discovery,huang2008efficient,achar2012unified,achar2013pattern}. However, the discovered frequent episode is still too simple and primitive. In some cases, FEM may lose some rich information, such as utility, important, risk, etc. Wu \textit{et al}. \cite{wu2013mining} presented the first attempt to solve the problem of high-utility episode mining (HUEM) in a complex event sequence. However, the proposed  UP-Span algorithm suffers from low efficiency in both runtime and memory consumption. Furthermore, the proposed upper-bound named Episode Weighted Utility (EWU) is a loose and basic utility bound for episodes. Guo \textit{et al}. then proposed the TSpan algorithm with several improvements for UP-Span in a much more efficient manner \cite{guo2014high}, which can save considerable search space and runtime. Then, Lin \textit{et al}. separately introduced some models to process complex event sequences and stock investment using high-utility episode mining and a genetic algorithm \cite{lin2015discovering,lin2017novel}. In addition, the top-$k$ issue of HUEM has been studied recently \cite{rathore2016top}.

\subsection{UPM in Big Data}

In the big data era, it requires more efficient frameworks of UPM to handle the big data issue. Several models are presented to address UPM in big data \cite{lin2015mining2,chen2016approximate,zihayat2016distributed}. Details are described below.

$\bullet$ \textbf{UPM in big itemset data}. PHUI-Growth (Parallel mining High-Utility Itemsets by pattern-Growth) \cite{lin2015mining2} is firstly proposed for parallel mining HUIs on Hadoop platform. It adopts the MapReduce \cite{dean2010mapreduce}  architecture to  partition the whole mining tasks. As a distributed parallel algorithm, PHUI-Miner with a sampling strategy is introduced by Chen \textit{et al.} \cite{chen2016approximate}. It extracts the approximate HUIs from big data. Recently, the study of parallel mining of top-k HUIs in Spark in-memory computing architecture is further proposed. It inherits several advantages of Spark \cite{zaharia2012resilient}.

$\bullet$ \textbf{UPM in big sequence data}. The BigHUSP model is the first work to discover distributed and parallel high-utility sequential patterns  \cite{zihayat2016distributed}. BigHUSP uses multiple steps of MapReduce \cite{dean2010mapreduce} to process big data in parallel. In contrast to the traditional HUSPM approaches, it can  deal with large-scale sequential data. MAHUSP \cite{zihayat2017memory} is a memory-adaptive approximation algorithm to efficiently discover high-utility sequential patterns over data streams. It employs a memory-adaptive mechanism using a bounded portion of memory, and guarantees that all HUSPs are discovered under certain circumstances. Experimental study shows that MAHUSP can not only discover HUSPs over data streams efficiently, but also adapt to memory allocation without sacrificing much of the quality of discovered HUSPs.

\subsection{UPM in Stream Data}

A data stream is an infinite sequence of data elements continuously arriving at a rapid rate \cite{golab2003issues,chi2004moment}. Mining useful patterns from data streams has become one of interesting problems of data mining \cite{manku2002approximate,li2008dsm,chi2004moment}. However, few works on mining data streams consider the utility factor embedded in data streams. Tseng \textit{et al}. first proposed the THUI-Mine (Temporal High-Utility Itemsets) model to mine temporal HUIs from data streams \cite{chu2008efficient}. THUI-Mine can effectively identify the temporal HUIs by generating fewer temporal 2-itemsets of HTWUIs. Thus, the execution time can be reduced significantly in mining all HUIs from data streams. In this way, the discovery process under all time windows of data streams can be achieved with limited memory space and less candidates. Then, researchers for HUIM proposed several stream mining models, such as MHUI-BIT (Mining High-Utility Itemsets based on BITvector) \cite{li2008fast}, MHUI-TID (Mining High-Utility Itemsets based on TIDlist) \cite{li2008fast}, and GUIDE (Generation of maximal high-Utility Itemsets from Data strEams) \cite{shie2010online,shie2012efficient}. GUIDE is a framework that mines the compact maximal HUIs from data streams with different models (i.e., the landmark, sliding, and time fading window models) \cite{shie2010online,shie2012efficient}. In \cite{ahmed2012interactive}, the HUS-tree (high-utility stream tree) and HUPMS algorithm (high-utility pattern mining over stream data) are proposed for incremental and interactive UPM over data streams with a sliding window.

\subsection{UPM with Various Interesting Constraints}

Up to now, most of the algorithms for UPM have been developed to improve the efficiency of the mining process, while effectiveness of the algorithms for UPM is also very important, because it is related to its usefulness for various data, constraints, and applications. Researchers in the field of utility-oriented pattern mining have proposed many algorithms and models to extend effectiveness. Many constraint-based UPM algorithms have been extensively developed for various problems, targeting a wide range of applications. For example, mining high-utility patterns with products' on-shelf time period \cite{lan2011discovery,lan2012thesis}, mining the up-to-date HUIs that reflect recent trends \cite{lin2015efficient1,gan2016mining}, mining discriminative high-utility patterns \cite{ahmed2011framework1,lin2017fdhup}, mining top-$k$ high-utility patterns without setting the minimum utility threshold \cite{wu2012mining,zihayat2014mining}, UPM with multiple minimum utility thresholds \cite{lin2016efficient1}, utility-based association rule mining \cite{lee2013utility,sahoo2015efficient}, UPM with consideration of various discount strategies \cite{lin2016fast1}, UPM by considering negative utility values \cite{lin2016fhn,gan2017mining}, UPM from uncertain data \cite{lin2016efficient2,lin2017efficiently}, and extracting non-redundant correlated HUIs \cite{gan2017extracting,gan2018coupm}. Obviously, UPM with various interesting constraints is an active research topic.

\subsection{Privacy Preserving for UPM}
Since more useful information is in the expected utility-based patterns than in that of the frequent itemsets or sequences, privacy preserving for high-utility pattern mining (PPUM) is more realistic and critical than privacy-preserving data mining (PPDM) \cite{agrawal2000privacy,lindell2000privacy,aggarwal2008general,zhu2017differentially}. Some preliminary studies have been done on this issue. Yeh \textit{et al}. first designed two models, named Hiding High-Utility Itemset First (HHUIF) and Maximum Sensitive Itemsets Conflict First (MSICF), to hide sensitive HUIs in PPUM \cite{yeh2010hhuif}. The main task of PPUM is to hide the sensitive high-utility itemsets (SHUIs). Lin \textit{et al}. first developed a genetic-algorithm-based method to hide the user-specified SHUIs by inserting the dummy transactions into the original databases \cite{lin2014ga}. Yun \textit{et al}. then developed a tree-based algorithm called the Fast Perturbation algorithm Using a Tree structure and Tables (FPUTT) for hiding SHUIs \cite{yun2015fast}. Then, other faster and more efficient algorithms were developed for PPUM, such as \cite{lin2016fast,lin2017efficient2}. A recent overview of PPUM has been reported by Gan \textit{at al}. \cite{gan2018privacy}.

\section{Open-Source Software and Datasets}
\label{sec:softwaredatasets}

\subsection{Open-Source Software}

Although the problem of UPM has been studied for more than 15 years, and the advanced topic of utility pattern mining also has been extended to many research fields, few implementations or source code of these algorithms have been released. This raises some barriers to other researchers in that they need to re-implement algorithms to use them or compare their performance with that of novel proposed algorithms. To make matters worse, this may introduce unfairness in running experimental comparisons, since the performance of pattern mining algorithms may commonly depend on the compiler and machine architecture used. We now list some open-source software specialized for UPM.

\textbf{$\bullet$ UP-Miner}. Tseng \textit{et al}. proposed a first-of-its-kind utility mining toolbox named Utility Pattern Miner (UP-Miner) \cite{tseng2015up}. UP-Miner provides various models for utility-oriented pattern mining. The main merits of UP-Miner have three aspects. First, to the best of our knowledge, it is the first-of-its-kind cross-platform utility mining system. Second, it provides complete Java implementations of 13 algorithms for discovering different types of utility-oriented patterns, such as high-utility itemset (HUI),  high-utility sequential rule (HUSR), high-utility sequential pattern (HUSP), and high-utility episode (HUE), as well as the concise representations of utility patterns. In addition, it offers four functionalities for processing utility-based databases. Third, the toolbox and relevant materials, including source codes, demo paper, benchmark datasets, and data generators, have been made public on Website\footnote{\url{http://bigdatalab.cs.nctu.edu.tw/software.php}} for the benefit of the research community.

\textbf{$\bullet$  SPMF}. As a well-known open-source data mining library, SPMF \cite{fournier2016spmf} offers implementations of many algorithms and has been cited in more than 700 research papers since 2010. SPMF is written in Java, and provides implementations of 170 data-mining algorithms, specializing in pattern mining. SPMF has the largest collection of implementations of various algorithms for pattern mining algorithms (i.e., FPM, ARM, SPM, etc.) and provides a user-friendly graphical interface\footnote{\url{http://www.philippe-fournier-viger.com/spmf/index.php}}. In particular, it also provides the relevant materials, including source codes, documentation, user instruction, benchmark datasets, data generators, and academic papers. SPMF offers up to 30 algorithms for high-utility pattern mining, such as Two-Phase, UP-Growth, UP-Growth+, HUI-Miner, d2HUP, EFIM, USpan, and many other state-of-the-art algorithms. More specifically, SPMF is distributed under the GPL v3 license and is suitable for both academic and industrial purposes.

\subsection{Datasets for UPM}

Several datasets are commonly used in the studies of UPM. All of them have been released at websites, such as SPMF \cite{fournier2016spmf}, UP-Miner \cite{tseng2015up}.

\textbf{Real datasets}. \textit{foodmart}: it is provided by Microsoft containing 21,556 customer transactions and 1,559 distinct items from an anonymous chain store. It contains the quantity and a unit profit of each item.  \textit{yoochoose-buys} commercial dataset was constructed  in the RecSys Challenge 2015\footnote{\url{https://recsys.acm.org/recsys15/challenge/}}. It contains a collection of 1,150,753 sessions from a retailer, where each session is encapsulating the click events. The total number of item IDs and category IDs is 54,287 and 347 correspondingly, with an interval of 6 months. \textit{UK-online}\footnote{\url{http://archive.ics.uci.edu/ml/datasets/Online+Retail/}}: it contains 541,909 transactions, which occurs between 01/12/2010 and 09/12/2011 for a UK-based and registered non-store online retail. The original data contains the real timestamp and many noise values. It has the attributes as InvoiceNo, StockCode,  Quantity, InvoiceDate, UnitPrice, CustomerID, etc.

\textbf{Semi-authentic datasets}. They are the real datasets\footnote{\url{http://fimi.ua.ac.be/data/}} (e.g., \textit{chess}, \textit{retail}, \textit{kosarak}, \textit{mushroom}, \textit{accidents}, \textit{BMSPOS2}) with synthetic utility values. The internal utility values are generated using a uniform distribution in [1, 10]. The external utility values are generated using a Gaussian (normal) distribution. Detailed description and characteristics of these real datasets can be referred to SPMF \cite{fournier2016spmf}, UP-Miner \cite{tseng2015up}, or existing UPM literature.

\textbf{Synthetic datasets}. There are some synthetic itemset-based or sequence-based datasets generated by IBM Quest Dataset Generator \cite{agrawal1994dataset}, which have been commonly used in UPM. For example, the itemset-based synthetic \textit{T10I4D100K}, \textit{T40I10D100K}; and the sequence synthetic \textit{C8S6T4I3D$|$X$|$K} are described in \cite{gan2019fast}.

%% file: 5.challenges.tex
\section{Open Challenges and Opportunities}
\label{sec:challenges}

Here, we discuss important open problems that have the potential to become future research areas in utility-oriented pattern mining. Owing to the rapid growth of the volume of data stored in databases, we have entered the era of Big Data. While analyzing utility-oriented patterns, we have identified numerous technical challenges and opportunities for UPM. We next highlight some important research opportunities, which are common to many, and sometimes all, UPM algorithms.

$\bullet$  \textbf{\textit{Application-driven algorithms}}. Up to now, most of the algorithms for UPM have been developed to improve the efficiency of mining process. The effectiveness of the algorithms for UPM is also very important, because it is related to the usefulness on various data, constraints, and applications. In general, as described in Section 2.3, the application-driven algorithms with many particular features of utility patterns reflect real-life problems of different applications in various fields. How to propose a specialized UPM model for different applications (e.g., business, web intelligent, risk perdition, smart city, financial analysis, Internet of Things, Biomedicine, smart transportation) and experimentally show its effectiveness is necessary and challenging. Moreover, the incorporation of domain knowledge \cite{cao2010domain} has a higher influence on performance for some data mining methods. Utility mining guided by domain knowledge thus provides many opportunities.

$\bullet$  \textbf{\textit{Developing more efficient algorithms}}. Traditionally, most pattern mining algorithms, especially UPM algorithms, are computationally expensive in terms of execution time and memory cost. This may be a serious problem for dense databases or databases containing numerous items/sequences or long transactions, depending on the minimum utility threshold chosen by the user. Although current UPM algorithms (e.g., HUI-Miner \cite{liu2012mining}, EFIM \cite{fournier2016efim}, and mHUIMiner \cite{peng2017mhuiminer}) are much efficient than previous Apriori-based and tree-based algorithms, there is still room for improvement. 1) It is important to reduce the search space, and this requires to design novel pruning strategies that rely on upper-bounds on the utility measure that are tighter than current measures. 2) Moreover, we can design novel data structures to more quickly calculate the utility and upper-bounds, and integrate constraints in the mining process to reduce the search space. 3) Fast approximate algorithms \cite{chen2016approximate} that guarantee a maximum error can also be developed.

$\bullet$     \textbf{\textit{Unified framework for UPM.}} Many variations of utility mining have been proposed to deal with various types of data and to solve different problems. The current paradigm used to solve utility-oriented pattern mining problem is to first define the definition of utility-based patterns with interest and their properties, and then develop an algorithm that can exploit the properties of the utility (e.g., upper bound) to efficiently mine them. Hence, this laborious process can be avoided if the following problem is solved: ``Is there a paradigm such that existing and new definitions of utility-based pattern (HUI \cite{liu2005two}, HUSP \cite{yin2012uspan,gan2019proum}, HUE \cite{wu2013mining}) can be solved by a unifying algorithm?" Owing to these challenges, the utility-oriented pattern mining problem, in its most general form, is not easy to solve. In fact, most of the existing utility mining techniques (e.g., HUIM \cite{liu2005two}, HUSPM \cite{yin2012uspan,gan2019proum}, HUEM \cite{wu2013mining}, etc.) solve a specific formulation of a specific problem. Therefore, how to formalize utility mining tasks in a generic framework is crucial and challenging. Focus on general principles and modeling of UPM rather than specific implementations is more important and challenging.

$\bullet$  \textbf{\textit{Deal with complex data}}. The amount of complex data has been explored during the past two decades, while most of the data mining and analysis approaches are not utility oriented. Many current techniques of UPM are not suited to dealing with various types of complex data, such as ``structured data"\footnote{\url{https://en.wikipedia.org/wiki/Structure_mining}} (i.e., pattern mining), ``unstructured data" \footnote{\url{https://en.wikipedia.org/wiki/Unstructured_data}} (including documents, health records, audio, video, images, etc.), and ``semi-structured data"\footnote{\url{https://en.wikipedia.org/wiki/Semi-structured_data}} (i.e., XML, JSON), and most of these are the heterogeneous data. More specifically, the dynamic data \cite{gan2018survey}, the uncertain data \cite{aggarwal2009frequent,lin2016efficient2}, the high-dimensional datasets of moderate size, or the very large datasets of moderate complexity in real-life applications are commonly seen in different domains and applications. Bridging this gap requires the solution of fundamentally new research problems, which can be grouped into the following  challenges: 1) how to define the utility function integrating with various rich features on complex data; 2) how to achieve utility maximization for the goal and mining task; and 3) how to develop new frameworks and algorithms to deal with new types of data. A need therefore arises for a better framework that extends the existing data mining methodologies, techniques, and tools, guided by utility and knowledge.

$\bullet$       \textbf{\textit{Large-scale data}}. Efficiently mining large-scale databases may result in a high computational cost and memory consumption. Under the batch model, traditional UPM algorithms must be repeatedly applied to obtain updated results when new data are inserted \cite{gan2018survey}. However, in the Big Data era, incrementally or dynamically processing data \cite{gan2018survey} and taking into account the results of prior analysis is crucial. There are some challenging research opportunities of UPM for handling large-scale data (as described in Section 4.7): how to design the parallelized UPM algorithms and how to develop the UPM algorithms based on the existing technologies of Big Data (i.e., MapReduce \cite{dean2010mapreduce} and Spark \cite{zaharia2012resilient}). Some other promising areas of research are the design of distributed, parallel, multi-core, or graphical-processing-unit (GPU)-based algorithms \cite{hong2011efficient,gan2017data} for UPM. There are some open challenges and opportunities to improve the scalability of utility mining tasks from resource-constraint devices to collaborative and hybrid execution models.

$\bullet$  \textbf{\textit{Scalable real-time pattern mining}}. There exists many interactive approaches for interactive data mining, but few have been extended to address the challenge of utility.  And it is not trivial to adapt them. One of the most important future challenges is to develop scalable high-utility pattern online mining approaches for streaming data from electronic commerce. Specifically, research should focus on algorithms that are sub-linear to the input or, at the very least, linear. Other computational challenges, such as the demands of the results being returned in real- or near-real-time, are the open issues in the data mining community. For example, real-time mining with optimization requires a new formalism and solving techniques. As mentioned before, increasing quantity and complexity of data demands scalable solutions. Using the existing computational infrastructures for real-time utility-oriented mining massive datasets may be a feasible way. 

%% file: 6.conclusion.tex
\section{Conclusions}
\label{sec:conclusion} 

The term utility is commonly used to mean ``the quality of being useful," and utilities are widely used in data-mining and decision-making processes to extract different useful kinds of knowledge. Utilities are subjective and can be acquired from domain experts/users. Utility mining in data is a vital task, with numerous high-impact applications, including cross-marketing, e-commerce, finance, medical, and biomedical applications. Up to now, many techniques and approaches have been extensively proposed for the task of UPM. In this survey, we have provided a comprehensive review of utility-oriented pattern mining, both in terms of current status and future directions. This survey describes various problems associated with mining utility-based patterns and methods for addressing these problems, including 1) \textit{high-utility itemset mining} (HUIM), 2) \textit{high-utility association rule mining} (HUARM), 3) \textit{high-utility sequential pattern mining} (HUSPM), 4) \textit{high-utility sequential rule mining} (HUSRM), and 5) \textit{high-utility episode mining} (HUEM). Overall, we have not only reviewed the most common, as well as the state-of-the-art, approaches for UPM but have also provided a comprehensive review of advanced UPM topics. Finally, we have identified several important issues and research opportunities for UPM.

\section{Acknowledgment}
We would like to thank the editors and anonymous reviewers for their detailed comments and constructive suggestions which have improved the quality of this paper. This research was partially supported by the China Scholarship Council Program.